%
% A/84475/PAP
%
% Nonequilibrium dynamics of a simple stochastic model
%
% by C Godreche and J M Luck
%
\magnification\magstep1
\hsize=16.5truecm
\vsize=22truecm
\baselineskip=14pt
\parindent=1em
\parskip=0.5ex plus 0.5ex
\font\ttlfnt=cmr10 scaled\magstep 2
\def\d{{\rm d}}
\def\dpar{{\partial}}
\def\e{{\rm e}}
\def\eps{\varepsilon}
\def\eff{_{\rm eff}}
\def\eq{_{\rm eq}}
\def\pl{_{\rm pl}}
\def\frac#1#2{{\displaystyle{\displaystyle#1\over\displaystyle#2}}}
\def\h{f^{\beta_1}}
\def\p#1#2{\frac{\partial#1}{\partial#2}}
\def\s{s}

\def\un{^{(1)}}
\def\deux{^{(2)}}
\def\t{\tau}
\def\L{\Lambda}
\def\LL{\L\eq}
\def\O#1{{{\cal O}\left(#1\right)}}
\def\C{{\cal C}}
\def\H{{\cal H}}
\def\K{{\cal K}}
\def\M{{\cal M}}
\def\W{{\cal W}}
\def\mid{\mathop{\vert}}
\def\fd{fluc\-tu\-a\-tion-dis\-si\-pa\-tion }

\noindent{\ttlfnt Nonequilibrium dynamics of a simple stochastic model}
\bigskip
\noindent C. Godr\`eche$^{1,2}$ and J.M. Luck$^{3}$
\bigskip
\noindent
$^{1}$ Service de Physique de l'\'Etat Condens\'e,
CEA-Saclay, 91191 Gif-sur-Yvette cedex, France

\noindent
$^{2}$ Groupe de Physique Statistique,
Universit\'e de Cergy-Pontoise, France

\noindent
$^{3}$ Service de Physique Th\'eorique,
CEA-Saclay, 91191 Gif-sur-Yvette cedex, France

\bigskip\null\bigskip
\noindent{\bf Abstract.}
We investigate the low-temperature dynamics of a simple stochastic model,
introduced recently in the context of the physics of glasses. The slowest
characteristic time at equilibrium diverges exponentially at low temperature.
On smaller time scales, the nonequilibrium dynamics
of the system exhibits an aging regime.
We present an analytical study of the scaling behaviour of the mean energy,
of its local correlation and response functions,
and of the associated \fd ratio
throughout the regime of low temperature and long times.
This analysis includes the aging regime, the convergence to equilibrium,
and the crossover behaviour between them.

\vfill
\hfill SPEC/97/030

\noindent To be published by Journal of Physics A
\hfill SPhT/97/050
\eject

\noindent{\bf 1 Introduction}
\vskip 12pt plus 2pt

The understanding of nonequilibrium phenomena owes much
to the analysis of simple models.
For example, kinetic Ising models play an important role in the study of
coarsening phenomena [1--3].
The so-called Backgammon model is another example of a simple model
which brings useful insight into the fields of nonequilibrium
statistical physics, slow dynamics and aging phenomena [4--6].
It was initially introduced in the context of the
physics of glasses,
as an illustration of the fact that a model possessing
entropic barriers only could reproduce some of the features of more complex
models possessing both energy and entropy barriers [7].
For example this model exhibits non-stationary dynamics,
with aging of two-times quantities [7--11]
and violation of the \fd theorem [12].

This work is devoted to the behaviour of the Backgammon model
at finite temperature.
We perform an exact analysis of the convergence of the mean
energy towards equilibrium, and of the relaxation of the fluctuations of the
energy in the equilibrium state.
We also give an analytical treatment of the nonequilibrium
dynamics of the model, which becomes asymptotically exact at long times and low
temperature.
This study is adapted from the theoretical framework introduced in ref. [11].
It allows us to show that the correlation and response functions
of the energy are aging at low temperature, to give their scaling forms,
and to compute the associated fluctuation-dissipation ratio.
This analysis is made possible by a specific feature of the model,
already present at equilibrium,
which is the existence of a clear-cut separation of
slow modes from fast ones,
respectively known as $\alpha$ and $\beta$ relaxation
in the phenomenology of glassy dynamics [6].
A parallel study of the dynamics of the density fluctuations in this model
will be the subject of a separate paper [13].

The model is defined as follows.
Consider $N$ particles,
initially distributed amongst $M$ boxes in a known fashion.
The energy of a box is equal to $0$ if the box is occupied
by one or more particles, and to $-1$ if it is empty.
The total energy of the system is the sum of the energies of each of the boxes;
it is thus equal to minus the number of empty boxes.
There is no reference to a spatial structure in the definition of the
model, hence it is of a mean-field type.
At each time step $1/N$,
a particle and a box are chosen independently at random.
The particle is moved to the box according to the Metropolis rule, i.e.,
with unit rate if the energy does not increase,
otherwise with rate $\e^{-\beta}$, where $\beta$ is the inverse temperature.
In particular, moves from a box containing one particle to an empty box
are always allowed, since they do not change the total energy.
This dynamics satisfies detailed balance.

The main features of this model are easy to grasp intuitively,
and can be explained in simple physical terms.
Consider for simplicity the zero-temperature dynamics [7--11].
The moves which decrease the energy consist in emptying
the boxes which contain only one particle, to the benefit of a non-empty box.
As time passes, the number of empty boxes is ever increasing,
and so does the mean occupancy of non-empty boxes.
It is therefore more and more rare to find a box with only one particle.
The dynamics is thus slowed down by purely entropic effects,
since there are no energy barriers.

Finally we recall that the present model is similar to a coarsening system
which never equilibrates [1, 2].
The mean occupancy of the non-empty boxes
plays the role of the mean domain size in the coarsening system.
At zero temperature it grows approximately as $\L(t)\sim\ln t$,
while the normalised two-time correlation function
of the energy has the scaling form [11]
$$
C(t,\s)\approx\left(\frac{\s}{t}\right)^{1/2}\frac{\ln\s}{\ln t}.
\eqno(1.1)
$$

\vskip 14pt plus 2pt
\noindent{\bf 2 Physical quantities and their dynamical evolution}
\vskip 12pt plus 2pt

Consider a finite system, made of $M$ boxes containing $N$ particles.
Let $N_i(t)$ be the occupation number of box number $i$ at time $t$,
i.e., the number of particles contained in that box.
We have
$$
\sum_{i=1}^M N_i(t)=N.
\eqno(2.1)
$$
The model is characterised by the following Hamiltonian $\H$, and action $S$,
at inverse temperature $\beta$:
$$
S=\beta\H=-\beta\sum_{i=1}^M\delta_{N_i,0}.
\eqno(2.2)
$$

We consider the system in the thermodynamic limit with a fixed density
($M,N\to\infty$, $\rho=N/M$ fixed).
In the following we will be interested in quantities
which involve the occupation number of one box only.
Therefore, taking box number 1 as a generic box,
we describe the dynamics of the system in terms of $N_1(t)$ only.
We denote by $f_k(t)$ the probability that box number 1
contains $k$ particles at time $t$:
$$
f_k(t)={\rm Prob}\big\{N_1(t)=k\big\}.
\eqno(2.3)
$$
Equivalently, $f_k(t)$ represents
the fraction of boxes containing $k$ particles at time $t$.

Restricting the analysis to the homogeneous
nonequilibrium initial condition where there is one particle per box,
we have $N_i(t=0)=1$ for all $i$, hence $\rho=1$ and
$$
f_k(0)=\delta_{k,1}.
\eqno(2.4)
$$

The occupation probabilities $f_k(t)$ obey the following dynamical equations,
whose derivation is given in Appendix A:
$$
\eqalign{
\frac{\d f_k(t)}{\d t}
&=\frac{k+1}{\L(t)}f_{k+1}(t)+f_{k-1}(t)-\left(1+\frac{k}{\L(t)}\right)f_k(t)
\qquad(k\ge2),\cr
\frac{\d f_1(t)}{\d t}
&=\frac{2}{\L(t)}f_2(t)+\mu(t)f_0(t)-2f_1(t),\cr
\frac{\d f_0(t)}{\d t}
&=f_1(t)-\mu(t)f_0(t),\cr
}
\eqno(2.5)
$$
where
$$
\eqalignno{
\frac{1}{\L(t)}&=1+(\e^{-\beta}-1)f_0(t),
&(2.6{\rm a})\cr
\mu(t)&=\e^{-\beta}+(1-\e^{-\beta})f_1(t).
&(2.6{\rm b})\cr
}
%\eqno(2.6)
$$

We introduce the compact notation
$$
\frac{\d f_k(t)}{\d t}=\sum_{\ell\ge0}\M_{k\ell}[f_0(t),f_1(t)]f_\ell(t),
\eqno(2.7)
$$
where the matrix $\M[f_0(t),f_1(t)]$ satisfies
$$
\sum_{k\ge0}\M_{k\ell}[f_0(t),f_1(t)]=0.
\eqno(2.8)
$$
The dynamical equations (2.5) are non-linear,
since the transition rates involve $\L(t)$ and $\mu(t)$,
and thus depend on $f_0(t)$ and $f_1(t)$.
We notice that the equation for $f_0(t)$
can be recast as $\d f_0(t)/\d t=f_1(t)/\L(t)-\e^{-\beta}f_0(t)$.
Finally, eqs. (2.5) ensure the conservation of the moments
$$
\sum_{k\ge 0}f_k(t)=1,\qquad
\langle N_1(t)\rangle=\sum_{k\ge 1}k\,f_k(t)=\rho=1.
\eqno(2.9)
$$

The dynamical equations (2.5) have a simple interpretation.
If $N_1(t)=k\ge 2$, the rate at which a particle enters box number 1 is unity,
while the rate {\it per particle} for leaving that box is $1/\L(t)$.
Hence the total departure rate is $k/\L(t)$.
Therefore, if $k<\L(t)$ the occupation number
of the box has a tendency to increase,
while if $k>\L(t)$ the tendency is opposite.
The occupation number $N_1(t)$ can also be viewed as the position at time
$t$ of a random walker on a semi-infinite chain,
taking values $N_1(t)=k=0,1,\ldots$, and obeying the master equation (2.5).
The walk is biased to the left if $k>\L(t)$, to the right if $k<\L(t)$,
so that the walker is confined around $k=\L(t)$ by a restoring force.
Hence $f_k(t)$, the distribution of $N_1(t)$,
is expected to be peaked around $k=\L(t)$ for long times, and for $\L(t)$
large enough, i.e., at low temperature.
Different rules hold at $k=0$ and 1.
In particular $k=0$ plays the role of an absorbing barrier [9].

Note that the average position of the walker $\langle N_1(t)\rangle$ is fixed
by eq. (2.9), so that the distribution $f_k(t)$ will in fact be bimodal at low
temperature.
This property will be illustrated by the case of the
equilibrium distribution (see Section 5).
Finally, there are two time scales in this system.
On the shorter time scale, $\L(t)$ may be considered as constant,
so that there is equilibration in a fixed confining potential.
On the longer scale, the potential itself evolves slowly, as $\L(t)$ increases.
These time scales will be identified in Sections 5 and 6
with those of $\alpha$ and $\beta$ relaxation, found in glassy dynamics.

\medskip
\noindent$\bullet$ {\it Mean energy}

The first quantity to be studied is the energy of the system.
According to eq. (2.2), the energy of box number $i$ is
$E_i(t)=-\delta_{N_i(t),0}$.
The mean energy per box of a thermodynamic system thus reads
$$
E(t)=\langle E_1(t)\rangle=-f_0(t).
\eqno(2.10)
$$
It can be expressed in terms of $\L(t)$ as
$$
E(t)=\frac{1-\L(t)}{(1-\e^{-\beta})\L(t)}.
\eqno(2.11)
$$
At low temperature, we have $E(t)\approx-1+1/\L(t)$
up to an exponentially small correction, proportional to $\e^{-\beta}$.
The mean occupancy of the non-empty boxes
is equal to $1/(1-f_0(t))\approx\L(t)$,
again up to an exponentially small correction at low temperature,
hence the interpretation of $\L(t)$ as a characteristic domain size,
in analogy with coarsening systems.

\medskip
\noindent$\bullet$ {\it Energy correlation function}

The correlation function $c(t,\s)$ between the values of the energy $E_1$
of box number 1 at the two times $\s$ (preparation time, or waiting time)
and $t$ (observation time) is defined as
$$
c(t,\s)
=\langle E_1(t)E_1(\s)\rangle-\langle E_1(t)\rangle\langle E_1(\s)\rangle,
\eqno(2.12)
$$
with $0\le\s\le t$.
Note that the present study is restricted to the local correlation function
$c(t,\s)$,
disregarding the non-diagonal correlations between two different boxes.

Since we have
$\langle E_1(t)E_1(\s)\rangle={\rm Prob}\big\{N_1(t)=0, N_1(\s)=0\big\}$,
we are led to introduce the joint probabilities
${\rm Prob}\big\{N_1(t)=k, N_1(\s)=0\big\}$.
Equivalently, we will consider the conditional probabilities $g_k(t,\s)$
that box number 1 contains $k$ particles at time $t$,
knowing that it was empty at the earlier time $\s$:
$$
g_k(t,\s)={\rm Prob}\big\{N_1(t)=k\mid N_1(\s)=0\big\}.
\eqno(2.13)
$$
These quantities obey the dynamical equations (see Appendix A)
$$
\p{g_k(t,\s)}{t}=\sum_{\ell\ge0}\M_{k\ell}[f_0(t),f_1(t)]g_\ell(t,\s),
\eqno(2.14)
$$
with the initial condition
$$
g_k(\s,\s)=\delta_{k,0}.
\eqno(2.15)
$$
Eq. (2.14) is now linear in the unknowns $g_k(t,\s)$.
It implies
$$
\sum_{k\ge0}g_k(t,\s)=1
\eqno(2.16)
$$
at all times $t\ge\s$.

The energy correlation function then reads
$$
c(t,\s)=f_0(\s)\big(g_0(t,\s)-f_0(t)\big).
\eqno(2.17)
$$
We also introduce the normalised correlation function
$$
C(t,\s)=\frac{c(t,\s)}{c(\s,\s)}=\frac{g_0(t,\s)-f_0(t)}{1-f_0(\s)},
\eqno(2.18)
$$
such that $C(\s,\s)=1$.

\medskip
\noindent$\bullet$ {\it Energy response function}

The local response function $r(t,\s)$ is a measure of
the change in the energy of box number 1 at time $t$, induced by an
infinitesimal change of the local temperature of the same box
at the earlier time $\s$.

More generally, assume that box number 1
is subjected to an arbitrary time-dependent inverse temperature $\beta_1(t)$,
while the other boxes ($i=2,\ldots,M$)
are subjected to a uniform and constant inverse temperature $\beta$.
To leading order at large $M$, the occupation probabilities of the boxes
$i=2,\ldots,M$ are still given by the $f_k(t)$,
while box number 1 has a different distribution,
depending on $\beta_1(t)$, denoted by $\h_k(t)$.

The dynamical equations for the $\h_k(t)$ are slightly different
from eqs. (2.5) (see Appendix A):
$$
\eqalign{
\frac{\d\h_k(t)}{\d t}
&=\frac{k+1}{\L(t)}\h_{k+1}(t)+\h_{k-1}(t)
-\left(1+\frac{k}{\L(t)}\right)\h_k(t)\qquad(k\ge2),\cr
\frac{\d\h_1(t)}{\d t}
&=\frac{2}{\L(t)}\h_2(t)+\mu_+(t)\h_0(t)-\left(1+\mu_-(t)\right)\h_1(t),\cr
\frac{\d\h_0(t)}{\d t}&=\mu_-(t)\h_1(t)-\mu_+(t)\h_0(t),\cr
}
\eqno(2.19)
$$
with the initial value
$$
\h_k(0)=\delta_{k,1}.
\eqno(2.20)
$$
In eqs. (2.19) we have set
$$
\eqalign{
\mu_+(t)&=\left(1-f_1(t)\right)\e^{-\beta_1(t)}
+f_1(t)\,\W(\beta_1(t)-\beta),\cr
\mu_-(t)&=1-f_0(t)+f_0(t)\,\W(\beta-\beta_1(t)),
}
\eqno(2.21)
$$
where
$$
\W(\Delta S)=\min(1,\e^{-\Delta S})
\eqno(2.22)
$$
is the Metropolis acceptance rate
associated with a change of action $\Delta S$ (see eq. (A.2)).

The response function is defined as
$$
r(t,\s)=-\left.\frac{\delta\langle E_1(t)\rangle}{\delta\beta_1(\s)}
\right|_{\,\beta_1(\s)=\beta}.
\eqno(2.23)
$$
Defining more generally
$$
h_k(t,\s)=\left.\frac{\delta\h_k(t)}{\delta\beta_1(\s)}
\right|_{\,\beta_1(\s)=\beta},
\eqno(2.24)
$$
we have
$$
r(t,\s)=h_0(t,\s).
\eqno(2.25)
$$

The dynamical equations for $h_k(t,\s)$ are obtained by taking the functional
derivative of eqs. (2.19) with respect to $\beta_1$ for $\beta_1(\s)=\beta$.
This yields
$$
\p{h_k(t,\s)}{t}=\sum_{\ell\ge0}\M_{k\ell}[f_0(t),f_1(t)]h_\ell(t,\s),
\eqno(2.26)
$$
with the initial condition
$$
h_k(\s,\s)=\left(\delta_{k,0}-\delta_{k,1}\right)\mu(\s)f_0(\s).
\eqno(2.27)
$$
Therefore
$$
\sum_{k\ge0}h_k(t,\s)=0
\eqno(2.28)
$$
at all times $t\ge\s$.
It is to be noted that the above equations are not affected by the fact
that the derivative of the Metropolis acceptance rate (2.22)
is not well-defined at $\Delta S=0$.
Indeed, $\d\W(\Delta S\to 0^+)/\d\Delta S=-1$ and
$\d\W(\Delta S\to 0^-)/\d\Delta S=0$ do not coincide.

We also introduce the normalised response function
$$
R(t,\s)=\frac{r(t,\s)}{r(\s,\s)}=\frac{h_0(t,\s)}{\mu(\s)f_0(\s)},
\eqno(2.29)
$$
such that $R(\s,\s)=1$.

\medskip
\noindent$\bullet$ {\it Fluctuation-dissipation ratio}

The \fd ratio $X(t,\s)$ provides a measure of the departure of the system
from equilibrium [6, 14, 12].
In the present case it is defined as
$$
r(t,\s)=X(t,\s)\p{c(t,\s)}{\s}.
\eqno(2.30)
$$
This definition contains no explicit temperature dependence,
in contrast with the \fd ratio of e.g. density fluctuations.

The energy \fd ratio can be expressed in terms of quantities introduced above.
Indeed $\dpar g_k(t,\s)/\dpar\s$ obeys the same dynamical equations as
$g_k(t,\s)$ or $h_k(t,\s)$ for $t>\s$,
since $\s$ only enters these equations as a parameter.
Moreover, let us integrate eq. (2.14) for the $g_k(t,\s)$ to first order
in $\eps=t-\s$:
$$
g_k(t=\s+\eps,\s)=\delta_{k,0}+\left(\delta_{k,1}-\delta_{k,0}\right)\mu(\s)
\eps+\O{\eps^2}.
\eqno(2.31)
$$
Hence
$$
\p{g_k(\s,\s)}{\s}=\left(\delta_{k,0}-\delta_{k,1}\right)\mu(\s).
\eqno(2.32)
$$
Comparing this last expression with the initial condition (2.27)
for the $h_k(t,\s)$ gives
$$
h_k(t,\s)=f_0(\s)\,\p{g_k(t,\s)}{\s}.
\eqno(2.33)
$$
As a consequence,
$$
\p{c(t,\s)}{\s}=h_0(t,\s)+\frac{\d f_0(\s)}{\d\s}\big(g_0(t,\s)-f_0(t)\big),
\eqno(2.34)
$$
so that the \fd ratio $X(t,\s)$ reads
$$
X(t,\s)=\frac{h_0(t,\s)}
{h_0(t,\s)+\frac{\d f_0(\s)}{\d\s}\big(g_0(t,\s)-f_0(t)\big)},
\eqno(2.35)
$$
or else
$$
X(t,\s)=\frac{r(t,\s)}{r(t,\s)+\frac{1}{E(\s)}\frac{\d E(\s)}{\d\s}\,c(t,\s)}.
\eqno(2.36)
$$
We can check from this expression that $X(t,\s)=1$ at equilibrium,
recovering thus the \fd theorem
$$
r(t,\s)=\p{c(t,\s)}{\s},
\eqno(2.37)
$$
while $0<X(t,\s)<1$ out of equilibrium.

\vskip 14pt plus 2pt
\noindent{\bf 3 Generating functions and integral representations}
\vskip 12pt plus 2pt

It is possible to obtain integral representations of the solutions of eqs.
(2.5), (2.14), (2.26) by means of the generating functions
$$
F(x,t)=\sum_{k\ge0}f_k(t)\,x^k,\quad
G(x,t,\s)=\sum_{k\ge0}g_k(t,\s)\,x^k,\quad
H(x,t,\s)=\sum_{k\ge0}h_k(t,\s)\,x^k.
\eqno(3.1)
$$
These functions obey the following partial differential equations
$$
\eqalignno{
\p{}{t}F(x,t)&=(x-1)\left(F(x,t)-{1\over\L(t)}\p{}{x}F(x,t)-Y_f(t)\right),
&(3.2{\rm a})\cr
\p{}{t}G(x,t,\s)&=
(x-1)\left(G(x,t,\s)-{1\over\L(t)}\p{}{x}G(x,t,\s)-Y_g(t,\s)\right),
&(3.2{\rm b})\cr
\p{}{t}H(x,t,\s)&=
(x-1)\left(H(x,t,\s)-{1\over\L(t)}\p{}{x}H(x,t,\s)-Y_h(t,\s)\right),
&(3.2{\rm c})\cr
}
%\eqno(3.2)
$$
where
$$
\eqalignno{
Y_f(t)&=(1-\e^{-\beta})f_0(t)=1-1/\L(t),
&(3.3{\rm a})\cr
Y_g(t,\s)&=\big(1-\mu(t)\big)g_0(t,\s)+\big(1-1/\L(t)\big)g_1(t,\s)&\cr
&=(1-\e^{-\beta})\big((1-f_1(t))g_0(t,\s)+f_0(t)g_1(t,\s)\big)
&(3.3{\rm b})\cr
&=(1-\e^{-\beta})\left(g_0(t,\s)+f_0(t)\p{g_0(t,\s)}{t}-
\frac{\d f_0(t)}{\d t}g_0(t,\s)\right);&\cr
}
%\eqno(3.3)
$$
Similar expressions for $Y_h(t,\s)$ are obtained by replacing $g$ by $h$
in the above three expressions for $Y_g$.
The initial conditions for eqs. (3.2) are derived from eqs. (2.4), (2.15),
(2.27):
$$
F(x,0)=x,\quad
G(x,\s,\s)=1,\quad
H(x,\s,\s)=(1-x)\mu(\s)f_0(\s),
\eqno(3.4)
$$
while the conservation of moments expressed by eqs. (2.9), (2.16), (2.28)
implies
$$
F(1,t)=\p{}{x}F(1,t)=1,\quad G(1,t,\s)=1,\quad H(1,t,\s)=0.
\eqno(3.5)
$$
Eqs. (3.2) can be formally solved by the method of characteristics
(see Appendix B), yielding
$$
\eqalignno{
F(x,t)&=\left(1+(x-1)\e^{-\t(t)}\right)\e^{(x-1)D(t,0)}
+\int_0^t\d u\,\K(x,t,u)Y_f(u),
&(3.6{\rm a})\cr
G(x,t,\s)&=\e^{(x-1)D(t,\s)}+\int_\s^t\d u\,\K(x,t,u)Y_g(u,\s),
&(3.6{\rm b})\cr
H(x,t,\s)&=\mu(\s)f_0(\s)\K(x,t,\s)+\int_\s^t\d u\,\K(x,t,u)Y_h(u,\s),
&(3.6{\rm c})\cr
}
%\eqno(3.6)
$$
with the definitions (B.4), (B.7):
$$
\eqalign{
\t(t)&=\int_0^t{\d u\over\L(u)},\cr
D(t,u)&=\int_u^t\d v\,\e^{\t(v)-\t(t)},\cr
\K(x,t,u)&=(1-x)\e^{\t(u)-\t(t)+(x-1)D(t,u)}.\cr
}
\eqno(3.7)
$$
Note that, in agreement with eq. (2.33), we have
$$
H(x,t,\s)=f_0(\s)\frac{\dpar G(x,t,\s)}{\dpar\s}.
\eqno(3.8)
$$

Setting $x=0$ in eqs. (3.6), we obtain
$$
\eqalignno{
f_0(t)&=\left(1-\e^{-\t(t)}\right)\e^{-D(t,0)}+\int_0^t\d u\,\K(0,t,u) Y_f(u),
&(3.9{\rm a})\cr
g_0(t,\s)&=\e^{-D(t,\s)}+\int_\s^t\d u\,\K(0,t,u)Y_g(u,\s),
&(3.9{\rm b})\cr
h_0(t,\s)&=\mu(\s)f_0(\s)\K(0,t,\s)+\int_\s^t\d u\,\K(0,t,u)Y_h(u,\s).
&(3.9{\rm c})\cr
}
%\eqno(3.9)
$$
Eq. (3.9a) is an implicit non-linear integral equation for $\L(t)$,
since $f_0(t)$, $\t(t)$, $D(t,u)$, and $\K(0,t,u)$
are defined in terms of $\L(t)$ by eqs. (2.6a), (3.7).
Let us finally point out the central role played by the kernel
$$
\K(0,t,u)=\e^{\t(u)-\t(t)-D(t,u)}=\frac{\dpar}{\dpar u}\e^{-D(t,u)}
\eqno(3.10)
$$
in the following sections.

\vfill\eject
%\vskip 14pt plus 2pt
\noindent{\bf 4 Infinite-temperature behaviour}
\vskip 12pt plus 2pt

At infinite temperature, the energy of configurations plays no role in the
dynamics.
The rate at which a particle enters any box is constant and equal to
unity, while the rate at which a particle leaves the box is equal to unity,
per particle.
Hence $\L(t)=\mu(t)=1$, and eqs. (3.6) provide explicit solutions.

\medskip
\noindent$\bullet$ {\it Mean energy}

Since $Y_f(t)=0$, eq. (3.6a) yields
$$
F(x,t)=\left(1+(x-1)\e^{-t}\right)\exp\left((x-1)(1-\e^{-t})\right).
\eqno(4.1)
$$
The occupation probabilities $f_k(t)$ thus read
$$
f_k(t)=\left((1-\e^{-t})^2+k\e^{-t}\right)
\frac{\left(1-\e^{-t}\right)^{k-1}\exp\left(\e^{-t}-1\right)}{k!}.
\eqno(4.2)
$$
In particular, the mean energy
$$
E(t)=-f_0(t)=(\e^{-t}-1)\exp\left(\e^{-t}-1\right)
=-\frac{1}{\e}+\frac{1}{2\e}\e^{-2t}+\cdots
\eqno(4.3)
$$
converges to its equilibrium value $E\eq=-\e^{-1}$
with a relaxation time $t\eq\un=1/2$.

\medskip
\noindent$\bullet$ {\it Energy correlation and response functions}

Similarly, noticing that $Y_g(t,\s)=Y_h(t,\s)=0$, we get
$$
\eqalign{
G(x,t,\s)&=\exp\left((x-1)(1-\e^{\s-t})\right),\cr
H(x,t,\s)&=(1-x)(1-\e^{-\s})
\exp\left(\e^{-\s}-1+\s-t+(x-1)(1-\e^{\s-t})\right).
}
\eqno(4.4)
$$
We thus have
$$
\eqalign{
g_k(t,\s)&=\frac{(1-\e^{\s-t})^k\exp(\e^{\s-t}-1)}{k!},\cr
h_k(t,\s)&=(1-\e^{-\s})\exp\left(\e^{-\s}+\e^{\s-t}-2+\s-t\right)
\frac{(1-\e^{\s-t})^{k-1}(1-k-\e^{\s-t})}{k!}.
}
\eqno(4.5)
$$
Note that the distributions (4.2), (4.5) thus found are all related to
Poisson distributions.
In particular the $g_k(t,\s)$ follow an exact Poisson law, with parameter
$1-\e^{\s-t}$.

The energy correlation and response functions then read
$$
\eqalign{
c(t,\s)&=(1-\e^{-\s})\exp\left(\e^{-\s}-2\right)
\big(\exp(\e^{\s-t})-(1-\e^{-t})\exp(\e^{-t})\big),\cr
r(t,\s)&=(1-\e^{-\s})\exp(\e^{-\s}+\e^{\s-t}-2+\s-t),
}
\eqno(4.6)
$$
so that the \fd ratio is given by
$$
X(t,\s)=\frac{(1-\e^{-\s})\e^{\s-t}}{(1-\e^{-\s})\e^{\s-t}
+\e^{-2\s}\big(1-(1-\e^{-t})\exp(\e^{-t}-\e^{\s-t})\big)}.
\eqno(4.7)
$$

The equilibrium correlation and response functions
only depend on the time difference $\theta=t-\s$:
$$
c\eq(\theta)=\e^{-2}\left(\exp(\e^{-\theta})-1\right),\qquad
r\eq(\theta)=\exp(\e^{-\theta}-\theta-2),
\eqno(4.8)
$$
and they obey the \fd theorem
$$
r\eq(\theta)=-\frac{\d c\eq(\theta)}{\d\theta}.
\eqno(4.9)
$$

The convergence of $c(t,s)$ and $r(t,s)$ to their equilibrium values
$c\eq(\theta)$ and $r\eq(\theta)$ is in
$\e^{-2\s}$, exhibiting again the relaxation time $t\eq\un=1/2$,
while the latter quantities fall-off as $\e^{-\theta}$,
with a relaxation time $t\eq\deux=1$.
The occurrence of these two different relaxation times
will be explained in Section 5.

\vskip 14pt plus 2pt
\noindent{\bf 5 Equilibrium properties and convergence to equilibrium}
\vskip 12pt plus 2pt

At finite temperature and for long enough times,
the model reaches a stationary state, corresponding to thermal equilibrium.
In this section, we analyse the
equilibrium behaviour of the quantities defined above
(mean energy, correlation and response functions).
In particular we compute the relaxation times
$t\eq\un$ and $t\eq\deux$ which characterise the convergence of the mean energy
to
equilibrium, and the relaxation of the local fluctuations of
energy at thermal equilibrium, respectively.

\medskip
\noindent$\bullet$ {\it Mean energy}

The stationary solution of the dynamical equations (2.5) reads
$$
\matrix{
(f_k)\eq=\e^{-\LL}\frac{\LL^{k-1}}{k!}\hfill&(k\ge 1),\cr
(f_0)\eq=\frac{\LL-1+\e^{-\LL}}{\LL}=\frac{\e^{\beta-\LL}}{\LL},&\cr
}
\eqno(5.1)
$$
where the value $\LL$ of $\L(t)$ at equilibrium is related to temperature by
$$
\e^\beta=1+(\LL-1)\e^{\LL}.
\eqno(5.2)
$$
The equilibrium thermodynamics of the model, summarised in Appendix C,
shows that $\LL$ is to be identified with the fugacity.

The occupation probabilities at equilibrium $(f_k)\eq$ for $k\ge 1$
are thus Poissonian with parameter $\LL$,
with the exception of $(f_0)\eq$, reflecting that empty boxes
are energetically favoured.
At low temperature, we have $\LL\approx\beta\gg 1$, so that
the $(f_k)\eq$ form a strongly bimodal distribution,
with many empty boxes, represented by the weight
$(f_0)\eq\approx1-1/\LL$ at $k=0$,
and few occupied boxes, represented by the weight $1/\LL$
concentrated in a narrow region of width $\sqrt{\LL}$ around $k=\LL$.

\medskip
\noindent$\bullet$ {\it Relaxation time of the mean energy}

The relaxation time $t\eq\un$ of the mean energy can be derived as follows.
First, using eq. (3.10), we recast eq. (3.9a) into the form
$$
f_0(t)=1-\K(0,t,0)-\int_0^t\d u\frac{\K(0,t,u)}{\L(u)}.
\eqno(5.3)
$$
Next, in order to linearise this expression around equilibrium, we set
$$
\L(t)=\LL+\delta\L(t).
\eqno(5.4)
$$
We have then, to first order in $\delta\L(t)$, setting $\eps=t-u$,
$$
\matrix{
f_0(t)=(f_0)\eq+\delta f_0(t),\hfill&
\t(t)=\frac{t}{\LL}+\delta\t(t),\hfill\cr
D(t,u)=\LL\big(1-\e^{-\eps/\LL}\big)+\delta D(t,u),\hfill&
\K(0,t,u)=K(\eps,\LL)+\delta\K(0,t,u),\hfill
}
\eqno(5.5)
$$
where
$$
\eqalign{
\delta f_0(t)&\approx\frac{\LL-1+\e^{-\LL}}{\LL^2(\LL-1)}\delta\L(t),\cr
\delta D(t,u)&\approx\frac{1}{\LL}
\int_0^\eps\d\zeta\big(\e^{-\zeta/\LL}-\e^{-\eps/\LL}\big)\delta\L(t-\zeta),
}
\eqno(5.6)
$$
and where the equilibrium kernel $K(\eps,\LL)=\lim_{t\to\infty}\K(0,t,t-\eps)$
reads
$$
K(\eps,\LL)=\exp\left(-\eps/\LL-\LL\big(1-\e^{-\eps/\LL}\big)\right).
\eqno(5.7)
$$
This expression has the following simple relationship
to the occupation probabilities $(f_k)\eq$:
$$
K(\eps,\LL)=\sum_{k\ge 1}k(f_k)\eq\,\e^{-k\eps/\LL}.
\eqno(5.8)
$$
For long enough times, the linearised form of eq. (5.3) reads
$$
\delta f_0(t)\approx\frac{1}{\LL^2}\int_0^t\d u\,K(t-u,\LL)\delta\L(u)
-\frac{1}{\LL}\int_0^t\d u\,\delta\K(0,t,u).
\eqno(5.9)
$$
Using eqs. (3.10), (5.6), we obtain
$$
\delta f_0(t)\approx\frac{1}{\LL^2}\int_0^\infty\d\eps
\left(K(\eps,\LL)-\e^{-\LL-\eps/\LL}\right)\delta\L(t-\eps).
\eqno(5.10)
$$
Taking the Laplace transform of eq. (5.10) leads to the characteristic equation
$$
\frac{\LL-1+\e^{-\LL}}{\LL-1}=\hat K(p,\LL)-\frac{\e^{-\LL}}{p+1/\LL},
\eqno(5.11)
$$
where the Laplace transform $\hat K(p,\LL)$ of $K(\eps,\LL)$ reads
$$
\eqalign{
\hat K(p,\LL)&=\int_0^\infty\d\eps\,\e^{-p\eps}\,K(\eps,\LL)
=\LL\int_0^1\d z\,z^{p\LL}\,\e^{\LL(z-1)}\cr
&=\e^{-\LL}\sum_{k\ge1}\frac{1}{p+k/\LL}\,\frac{\LL^{k-1}}{(k-1)!}.
}
\eqno(5.12)
$$
Notice that the subtracted term in the right-hand side of eq. (5.11)
is just the first term $(k=1)$ in the last expansion of $\hat K(p,\LL)$.

Eq. (5.11) has an infinite sequence of real negative solutions,
which we denote by $p=-p\un_k$, with $k\ge2$.
The characteristic values $p\un_k$ represent the
inverse characteristic times of the relaxation of energy
to its equilibrium value.
In particular,
the relaxation time $t\eq\un$ is the inverse of the smallest characteristic
value:
$$
t\eq\un=\frac{1}{p\un_2}.
\eqno(5.13)
$$

At infinite temperature $(\LL=1)$, the characteristic values are the
poles of $\hat K(p,\LL)$ (except the first one, because of the subtraction),
hence $p\un_k=k$ with $k\ge2$, and $t\eq\un=1/2$.
As temperature decreases,
the spectrum of characteristic values is continuously deformed,
and the solutions of eq. (5.11) stay within a bounded distance from
the poles of $\hat K(p,\LL)$, namely $p\un_k\approx-k/\LL$.
At low temperature, the relaxation time $t\eq\un$ is exponentially large in
$\LL$.
Indeed, we have the expansion around $p=0$
$$
\hat K(p,\LL)=1-\e^{-\LL}-p\LL\e^{-\LL}I(\LL)+\cdots,
\eqno(5.14)
$$
where [11]
$$
I(\L)
=\int_0^1{\d z\over z}(\e^{\L z}-1)
=\sum_{n\ge1}{\L^n\over n\,n!}
\approx{\e^\L\over\L}\sum_{\ell\ge 0}{\ell!\over\L^\ell}.
\eqno(5.15)
$$
By inserting the expansion (5.14) into eq. (5.11), we obtain
$$
t\eq\un\approx\frac{\LL-1}{\LL}I(\LL),
\eqno(5.16)
$$
i.e.,
$$
t\eq\un\approx\frac{\e^{\LL}}{\LL}\left(1+\frac{1}{\LL^2}+\cdots\right)
\approx\frac{\e^{\beta}}{\beta^2}\left(1+\frac{2\ln\beta+1}{\beta}
%+\frac{3\ln^2\beta+\ln\beta}{\beta^2}
+\cdots\right),
\eqno(5.17)
$$
while all the other characteristic times are of order $\LL\approx\beta$
at low temperature.

\medskip
\noindent$\bullet$ {\it Energy correlation and response functions:
generalities}

At equilibrium, namely for $\s\gg t\eq\un$, the energy correlation and response
functions become stationary, i.e., invariant under time translation.
They only depend on the time difference $\theta=t-\s$:
$$
\eqalignno{
c\eq(\theta)&=(f_0)\eq\big((g_0)\eq(\theta)-(f_0)\eq\big),
&(5.18{\rm a})\cr
r\eq(\theta)&=(h_0)\eq(\theta).
&(5.18{\rm b})\cr
}
%\eqno(5.18)
$$

Their expressions can be derived from eqs. (3.9b, c),
which respectively become, setting $\eps=t-u$ in the integrals,
$$
\eqalignno{
(g_0)\eq(\theta)&=\e^{-\LL\left(1-\e^{-\theta/\LL}\right)}
+\int_0^\theta\d\eps\,K(\eps,\LL)(Y_g)\eq(\theta-\eps),
&(5.19{\rm a})\cr
(h_0)\eq(\theta)&=\e^{-\LL}\,K(\theta,\LL)
+\int_0^\theta\d\eps\,K(\eps,\LL)(Y_h)\eq(\theta-\eps),
&(5.19{\rm b})\cr
}
%\eqno(5.19)
$$
where the equilibrium kernel $K(\eps,\LL)$ has been defined in eq. (5.7),
and where
$$
(Y_g)\eq(\theta)=(1-\e^{-\beta})
\Big(\big(1-(f_1)\eq\big)(g_0)\eq(\theta)+(f_0)\eq(g_1)\eq(\theta)\Big),
\eqno(5.20)
$$
and a similar expression for $(Y_h)\eq$.
First of all, we have the identity
$$
(h_0)\eq(\theta)=-(f_0)\eq{\d(g_0)\eq(\theta)\over\d\theta}
\eqno(5.21)
$$
(see eq. (2.33)), which expresses the \fd theorem (2.37) in the form
$$
r\eq(\theta)=-\frac{\d c\eq(\theta)}{\d\theta},
\eqno(5.22)
$$
generalising eq. (4.9) to any finite temperature.

The integral equations (5.19) can be solved by Laplace transforms.
Indeed, using the definition of $Y_h$ and the dynamical equation (2.26)
for $h_0$, we obtain
$$
(\hat h_0)\eq(p)=(f_0)\eq\big(1-p(\hat g_0)\eq(p)\big)
=\frac{\e^{-\LL}\hat K(p,\LL)}
{\LL-(\LL-1)\left(p+\frac{1}{(f_0)\eq}\right)\hat K(p,\LL)}.
\eqno(5.23)
$$
The function in the rightmost side of eq. (5.23) is a meromorphic function,
with an infinite sequence of poles, situated on the real negative axis,
which we denote by $p=-p\deux_k$, with $k\ge 1$.
The $p\deux_k$ give the inverse characteristic times
of the energy correlation and response functions.
In particular the relaxation time of these functions
is the inverse of the smallest pole:
$$
t\eq\deux=\frac{1}{p\deux_1}.
\eqno(5.24)
$$
We have therefore
$$
\eqalignno{
r\eq(\theta)&=(h_0)\eq(\theta)=\sum_{k\ge1}a_k\e^{-p\deux_k\theta},
&(5.25{\rm a})\cr
c\eq(\theta)&=(f_0)\eq\big((g_0)\eq(\theta)-(f_0)\eq\big)
=\sum_{k\ge1}\frac{a_k}{p\deux_k}\e^{-p\deux_k\theta},
&(5.25{\rm b})\cr
}
%\eqno(5.25)
$$
where the $a_k$ are the residues of the rightmost side of eq. (5.23)
at the poles $p=-p\deux_k$.
The values of these functions at $\theta=0$ read
$$
\eqalignno{
r\eq(0)&=\sum_{k\ge1}a_k=\e^{-\LL},
&(5.26{\rm a})\cr
c\eq(0)&=\sum_{k\ge1}\frac{a_k}{p\deux_k}=(f_0)\eq(1-(f_0)\eq)
=\frac{\left(1-\e^{-\LL}\right)\left(\LL-1+\e^{-\LL}\right)}{\LL^2}.
&(5.26{\rm b})\cr
}
%\eqno(5.26)
$$

At infinite temperature $(\LL=1)$,
the poles of eq. (5.23) coincide with those of $\hat K(p,\LL)$,
hence $p\deux_k=k$ for $k\ge1$, and $t\eq\deux=1$.
Again, as temperature decreases, the whole spectrum is continuously deformed,
and the poles stay close to those of $\hat K(p,\LL)$: $p\deux_k\approx-k/\LL$.
At low temperature, the relaxation time $t\eq\deux$ is exponentially divergent,
while all the other characteristic times remain of order $\LL$.
Indeed, by expanding the denominator of the rightmost side of eq. (5.23)
to first order in $p$, and using again the expansion (5.14), we obtain
$$
t\eq\deux
\approx\frac{(\LL-1)\e^{\LL}}{\LL^2}\big(\LL^2\e^{-\LL}I(\LL)+1-\LL\big),
\eqno(5.27)
$$
i.e.,
$$
t\eq\deux\approx\frac{2\e^{\LL}}{\LL}\left(1+\frac{2}{\LL^2}+\cdots\right)
\approx\frac{2\e^{\beta}}{\beta^2}\left(1+\frac{2\ln\beta+1}{\beta}
%+\frac{3\ln^2\beta+\ln\beta+1}{\beta^2}
+\cdots\right).
\eqno(5.28)
$$
Let us remark that the relaxation time $t\eq\deux$ of the energy correlation
and
response functions is roughly twice as large as the relaxation time $t\eq\un$
of the mean energy.
Indeed the ratio $t\eq\deux/t\eq\un$ is equal to two,
both at infinite temperature $(t\eq\un=1/2,$ $t\eq\deux=1)$ and at zero
temperature.
We have in the low-temperature regime
$$
\frac{t\eq\deux}{t\eq\un}\approx2\left(1+\frac{1}{\LL^2}+\cdots\right)
\approx2\left(1+\frac{1}{\beta^2}+\cdots\right).
\eqno(5.29)
$$
This ratio takes its maximal value 2.20359 for $\beta=4.1986$.
Figure 1 shows the dependence on temperature of
$t\eq\un$ and $t\eq\deux$, together with their low-temperature estimates
(5.17), (5.28).

We have thus met two spectra of inverse characteristic times, the $p\un_k$,
associated with the convergence of the mean energy to its equilibrium value,
and the $p\deux_k$, associated with the energy correlation
and response functions at equilibrium.
These spectra have a very similar dependence on temperature.
Their low-temperature behaviour is characterised as follows:
the relaxation times $t\eq\un$ and $t\eq\deux$ become exponentially large
in $\LL$ or in $\beta$,
while the subleading characteristic times remain of order $\LL$.
In analogy with the phenomenology of glassy dynamics,
the slow modes corresponding to the time scale
$\theta\sim t\eq\un\sim t\eq\deux\sim\e^{\LL}$
will be referred to as $\alpha$ relaxation,
while the fast modes corresponding to the microscopic time scale $\theta\sim 1$
or $\theta\sim\LL$ will be referred to as $\beta$ relaxation.
This separation of the $\alpha$ and $\beta$ relaxation processes
in a very clear-cut way is a remarkable feature of the present model.

\medskip
\noindent$\bullet$ {\it Energy correlation function}

Let us now consider the low-temperature behaviour of the
equilibrium correlation function $c\eq(\theta)$.
Using again the expansion (5.14), we estimate the residue $a_1$ as
$$
a_1\approx\frac{\LL-1}{\LL^2}p\deux_1
\approx\frac{\e^{-\LL}}{\LL^2\e^{-\LL}I(\LL)+1-\LL}
\approx\frac{\e^{-\LL}}{2}\left(1-\frac{1}{\LL}-\frac{2}{\LL^2}+\cdots\right).
\eqno(5.30)
$$
This expression shows that the sum of eq. (5.26b)
is overwhelmingly dominated by its first term $(k=1)$,
since the relative difference between the full sum and the first term
is exponentially small.
As a consequence, the energy correlation function at equilibrium
is very close to being a pure decaying exponential,
corresponding to $\alpha$ relaxation:
$$
C\eq(\theta)\approx\e^{-\theta/t\eq\deux},
\eqno(5.31)
$$
with exponentially small corrections at low temperature.

\medskip
\noindent$\bullet$ {\it Energy response function}

In contrast with the simple behaviour (5.31) of the correlation function,
the response function $r\eq(\theta)$ exhibits both $\alpha$ and $\beta$
relaxation processes at low temperature.

\noindent ---
In the regime of $\beta$ relaxation $(\theta\sim 1)$,
the response function at low temperature is obtained
by expanding eq. (5.12) for $\LL$ large but $p$ finite:
$$
\hat K(p,\LL)=\frac{1}{p+1}-\frac{p}{(p+1)^3\LL}
%+\frac{p(p-2)}{(p+1)^5\LL^2}
+\cdots,
\eqno(5.32)
$$
hence
$$
\hat h_0(p)\approx\e^{-\LL}\frac{p+1}{p(p+2)},
\eqno(5.33)
$$
and
$$
R\eq(\theta)\approx\frac{1}{2}(1+\e^{-2\theta}),
\eqno(5.34)
$$
up to correction terms of order $1/\LL$.

\noindent ---
In the regime of $\alpha$ relaxation $(\theta\sim t\eq\deux)$,
the asymptotic fall-off of the response function reads
$$
r\eq(\theta)\approx a_1\e^{-\theta/t\eq\deux},\qquad
R\eq(\theta)\approx a_1\e^{\LL}\e^{-\theta/t\eq\deux},
\eqno(5.35)
$$
where $a_1$ has been estimated in eq. (5.30).

\noindent ---
In the crossover regime $(1\ll\theta\ll t\eq\deux)$,
the normalised response function $R\eq(\theta)$
exhibits a non-trivial $\beta$-to-$\alpha$ plateau value
$$
R\pl\approx a_1\e^{\LL}\approx\frac{1}{2}\left(1-\frac{1}{\LL}
%-\frac{2}{\LL^2}
+\cdots\right)
\approx\frac{1}{2}\left(1-\frac{1}{\beta}
%-\frac{\ln\beta+2}{\beta^2}
+\cdots\right),
\eqno(5.36)
$$
asymptotically equal to $1/2$ at low temperature.

Figure 2 shows log-log plots of
(a) the normalised correlation function $C\eq(\theta)$
and (b) the normalised response function $R\eq(\theta)$,
for various values of the inverse temperature $\beta$.
The different properties of these two functions,
discussed above, are clearly apparent.

\vskip 14pt plus 2pt
\noindent{\bf 6 Nonequilibrium behaviour at low temperature}
\vskip 12pt plus 2pt

The equilibrium analysis of Section 5 shows that, at low temperature, the model
exhibits slow modes ($\alpha$ relaxation, characterised by the time scales
$t\eq\un$ and $t\eq\deux$), well separated from fast ones ($\beta$ relaxation,
characterised by the microscopic time scale $\sim 1$ or $\LL$).
In this section, we show that this separation of modes at low temperature still
holds out of equilibrium.
This property is at the basis of the analysis presented below.
The latter consists in an adiabatic elimination of the fast modes
(nonequilibrium $\beta$ relaxation),
generalising the method used in ref. [11].
We give an analytical treatment of the
quantities of interest (mean energy, correlation and response functions),
which becomes asymptotically exact for long times ($\s$ and $t-\s\gg 1$,
irrespectively of the relative positions of these variables
with respect to the relaxation times $t\eq\un$ and $t\eq\deux$)
and low temperatures.
This regime, hereafter called the (nonequilibrium) $\alpha$ regime,
encompasses in particular the aging regime
$(s$ and $t-\s\ll t\eq\un,t\eq\deux)$,
and the regime of convergence to equilibrium $(\s\gg t\eq\un,t\eq\deux)$.

We begin the analysis by considering the difference
$$
\t(t)-\t(u)=\int_u^t{\d v\over\L(v)},
\eqno(6.1)
$$
for $t\gg 1$ but $\eps=t-u\sim 1$.
Since $\L(t)$ is varying slowly in the $\alpha$ regime under consideration,
we have $\d\L(t)/\d t\ll\L(t)$.
Hence it is justified to perform a Taylor expansion of $\L(v)$ around $v=t$:
$$
\t(t)-\t(u)=\int_u^t\d v
\left(\frac{1}{\L(t)}+(t-v)\frac{\d\L(t)/\d t}{\L(t)^2}+\cdots\right)
=\frac{\eps}{\L(t)}+\frac{\eps^2}{2}\frac{\d\L(t)/\d t}{\L(t)^2}+\cdots
\eqno(6.2)
$$
This expansion makes sense when $\eps$
is small with respect to the characteristic time of variation of $\L(t)$.
Note that performing a Taylor expansion around the lower bound $v=u$
would lead to a result equivalent to eq. (6.2).
We have kept the first correction term,
in order to show explicitly that it is of relative order
$(\d\L(t)/\d t)/\L(t)\ll 1$.
Indeed this quantity will turn out to be exponentially small
in $\L(t)$ throughout the $\alpha$ regime.

The result (6.2) can be used to derive the following estimates
$$
\eqalign{
D(t,u)&\approx\L(t)\big(1-\e^{-\eps/\L(t)}\big),\cr
\K(0,t,u)&\approx\exp
\left(-\eps/\L(t)-\L(t)\big(1-\e^{-\eps/\L(t)}\big)\right)=K(\eps,\L(t)),
}
\eqno(6.3)
$$
where $K(\eps,\L(t))$ is obtained by replacing $\LL$ by $\L(t)$
in the expressions (5.12).
As mentioned above,
the approximate results (6.2), (6.3) hold for $t\gg 1$ but $\eps=t-u\sim 1$.
It will, however, be legitimate to employ them
for an arbitrary time difference $\eps$,
since the quantities $D(t,u)-\L(t)$ and $\K(0,t,u)$
decay exponentially with $\eps$ anyhow.

\medskip
\noindent$\bullet$ {\it Mean energy}

Considering now the integral equation (3.9a) for $\L(t)$,
we perform a Taylor expansion of $Y_f(u)$ near $u=t$.
It can be checked a posteriori that
it is consistent to keep the first two terms of this expansion.
Neglecting furthermore quantities which are exponentially small in $t$,
we transform eq. (3.9a) into the differential equation
$$
f_0(t)\approx\e^{-\L(t)}+J_0(t)Y_f(t)+J_1(t)\frac{\d Y_f(t)}{\d t}+\cdots,
\eqno(6.4)
$$
with
$$
\eqalign{
J_0(t)&=\int_0^t\d u\,\K(0,t,u)\cr
&\approx\int_0^\infty\d\eps\,K(\eps,\L(t))
=\hat K(0,\L(t))=1-\e^{-\L(t)},\cr
J_1(t)&=\int_0^t\d u\,(u-t)\K(0,t,u)\cr
&\approx-\int_0^\infty\d\eps\,\eps\,K(\eps,\L(t))
=\left.\frac{\d\hat K(p,\L(t))}{\d p}\right|_{p=0}
=-\L(t)\e^{-\L(t)}I(\L(t)).
}
\eqno(6.5)
$$

By inserting the estimates (6.5) into eq. (6.4), using eqs. (2.6a), (3.3a),
and consistently neglecting terms of relative order
$\e^{-\L(t)}$ or $\e^{-\beta}$, we obtain
$$
\frac{\d\L(t)}{\d t}\approx A(\L(t),\LL),
\eqno(6.6)
$$
with
$$
A(\L,\LL)=\frac{1}{I(\L)}\left(1-\frac{(\L-1)\e^{\L}}{(\LL-1)\e^{\LL}}\right).
\eqno(6.7)
$$
The evolution equation (6.6) for $\L(t)$ describes the relaxation of energy
throughout the $\alpha$ regime.
Let us consider the following cases:

\noindent ---
At zero temperature, and more generally in the aging regime $(t\ll t\eq\un)$,
we recover the result [11]:
$$
\frac{\d\L}{\d t}\approx\frac{1}{I(\L)},
\eqno(6.8)
$$
hence
$$
t-t_0\approx\int_0^\L\d\L'\,I(\L')
=\int_0^1{\d z\over z^2}(\e^{\L z}-1-\L z)
=\sum_{n\ge1}{\L^{n+1}\over n(n+1)!}
\approx{\e^\L\over\L}\sum_{\ell\ge 0}{(\ell+1)!\over\L^\ell},
\eqno(6.9)
$$
or else
$$
\L(t)\approx\ln t+\ln\ln t+\frac{\ln\ln t-2}{\ln t}+\cdots
\eqno(6.10)
$$
It is natural to choose the constant $t_0$ in order to fulfil
the initial condition $\L(t=0)=1$, yielding $t_0=-0.59962$.
Anyhow, $t_0$ only brings corrections to the scaling law (6.10)
which are exponentially small in $\L(t)$,
i.e., comparable to the corrections terms which have been neglected
during the analysis.

\noindent ---
At finite (low) temperature, thermal equilibrium
is reached for long enough times.
The convergence of $\L(t)$ towards the equilibrium value $\LL$ is exponential:
$$
\LL-\L(t)\approx a(\LL)\e^{-t/t\eq\un}.
\eqno(6.11)
$$
The relaxation time $t\eq\un$ reads
$$
t\eq\un\approx-\frac{1}{\dpar A(\L,\LL)/\dpar\L\big\vert_{\L=\LL}}
\approx\frac{\LL-1}{\LL}I(\LL),
\eqno(6.12)
$$
in agreement with the result (5.16) at equilibrium,
while the prefactor $a(\LL)$ is given by
$$
t\eq\un\ln\frac{a(\LL)}{\LL}=\int_0^{\LL}\d\L
\left(\frac{1}{A(\L,\LL)}-\frac{t\eq\un}{\LL-\L}\right),
\eqno(6.13)
$$
yielding, after some algebra, $a(\LL)\approx 1$ for $\LL$ large.

\noindent ---
When time $t$ is comparable to the relaxation time $t\eq\un$,
$\L(t)$ changes over from the zero-temperature logarithmic growth (6.10)
to the exponential convergence (6.11) towards the equilibrium value $\LL$.
The scaling behaviour throughout this crossover
can be described, to leading order in $1/\L$, by approximating eq. (6.7) as
$$
A(\L,\LL)\approx\L\big(\e^{-\L}-\e^{-\LL}\big).
\eqno(6.14)
$$
This equation can be integrated, to leading order in $1/\L(t)$ and $1/\LL$.
Using the definition (6.12), we are left with the explicit
interpolation formula
$$
\frac{t}{t\eq\un}\approx-\frac{\LL}{\L(t)}\ln\left(1-\e^{\L(t)-\LL}\right).
\eqno(6.15)
$$

Figure 3 illustrates the accuracy of the low-temperature analysis
in the $\alpha$ regime.
Exact values for $\L(t)$,
obtained by numerical integration of the dynamical equations (2.5),
are found to nicely agree with the prediction of the evolution equation (6.6),
for various (large) values of the inverse temperature $\beta$.

\medskip
\noindent$\bullet$ {\it Energy correlation and response functions:
generalities}

In analogy with the derivation of eq. (6.4),
we transform eqs. (3.9b, c) into the following differential equations
$$
\eqalign{
g_0(t,\s)&\approx\e^{-\L(t)}+
L_0(t,\s)Y_g(t,\s)+L_1(t,\s)\p{Y_g(t,\s)}{t}+\cdots,\cr
h_0(t,\s)&\approx
L_0(t,\s)Y_h(t,\s)+L_1(t,\s)\p{Y_h(t,\s)}{t}+\cdots,\cr
}
\eqno(6.16)
$$
with
$$
L_0(t,\s)=\int_\s^t\d u\,\K(0,t,u),\qquad
L_1(t,\s)=\int_\s^t\d u\,(u-t)\K(0,t,u).
\eqno(6.17)
$$
These integrals differ from $J_0(t)$ and $J_1(t)$ of eq. (6.4)
only through the values of their lower bounds.
They are therefore given by the expressions (6.5),
again up to exponentially small corrections.

By inserting the estimates (6.5) into eq. (6.16),
using eq. (3.3b) and similar expressions for $Y_h$,
and again consistently neglecting terms of relative order
$\e^{-\L}$ or $\e^{-\beta}$, we obtain the homogeneous differential equations
$$
\frac{1}{c(t,\s)}\frac{\dpar c(t,\s)}{\dpar t}
\approx\frac{1}{r(t,\s)}\frac{\dpar r(t,\s)}{\dpar t}
\approx-B(\L(t),\LL),
\eqno(6.18)
$$
with
$$
B(\L,\LL)=\frac
{A(\L,\LL)+\L^2\left(\e^{-\L}+\frac{\e^{-\LL}}{\LL-1}\right)}
{\L\left(\L^2\e^{-\L}I(\L)+1-\L\right)}.
\eqno(6.19)
$$
Changing the time variables from $\s$ and $t$ to $\L(\s)$ and $\L(t)$,
we have the alternative form
$$
\frac{1}{c(t,\s)}\frac{\dpar c(t,\s)}{\dpar\L(t)}
\approx\frac{1}{r(t,\s)}\frac{\dpar r(t,\s)}{\dpar\L(t)}
\approx-\alpha(\L(t),\LL),
\eqno(6.20)
$$
with
$$
\alpha(\L,\LL)=\frac{B(\L,\LL)}{A(\L,\LL)}.
\eqno(6.21)
$$

These evolution equations describe the nonequilibrium
relaxation of the correlation and response functions of the energy
throughout the $\alpha$ regime at low temperature,
as we shall see in more detail.

\medskip
\noindent$\bullet$ {\it Energy correlation function}

The correlation function $c(t,\s)$ can be directly estimated from eq. (6.20)
with the initial value $c(\s,\s)=f_0(\s)(1-f_0(\s))$, or $C(\s,\s)=1$.
Indeed, we know from the analysis of the equilibrium properties
that this function varies very little in the $\beta$ regime $(t-\s\sim 1)$.
We thus obtain
$$
C(t,\s)\approx\exp\left(-\int_{\L(\s)}^{\L(t)}\d\L'\,\alpha(\L',\LL)\right).
\eqno(6.22)
$$
This result can be recast into a multiplicative scaling law:
$$
C(t,\s)\approx\frac{\Phi(\L(\s),\LL)}{\Phi(\L(t),\LL)},
\eqno(6.23)
$$
with
$$
\Phi(\L,\LL)=\exp\left(\int_{\L_0}^\L\d\L'\,\alpha(\L',\LL)\right).
\eqno(6.24)
$$
An interesting interpretation of eq. (6.23)
consists in introducing an effective waiting time
$$
\s\eff=\frac{\Phi(\L(\s),\LL)}{\d\Phi(\L(\s),\LL)/\d\s}
=\frac{1}{B(\L(s),\LL)},
\eqno(6.25)
$$
along the lines of ref. [6].
The correlation function and the effective waiting time
exhibit several regimes:

\noindent ---
At zero temperature, and more generally in the aging regime
$(\s\ll t\eq\deux)$, we have
$$
\alpha(\L,\infty)
=\frac{\L^2\e^{-\L}I(\L)+1}{\L\left(\L^2\e^{-\L}I(\L)+1-\L\right)}
\approx\frac{1}{2}\left(1+\frac{1}{\L}-\frac{2}{\L^2}+\cdots\right),
\eqno(6.26)
$$
hence the approximately square-root behaviour
$$
\Phi(\L(\s),\infty)\approx c_0\left(\L(\s)\e^{\L(\s)}\right)^{1/2}
\left(1-\frac{1}{\L(\s)}+\cdots\right)
\approx c_0\s^{1/2}\ln\s\left(1+\frac{\ln\ln\s-2}{\ln\s}+\cdots\right).
\eqno(6.27)
$$
The absolute prefactor $c_0$ depends on the initial conditions, according to
$$
c_0=\exp\left(\int_{\L_0}^\infty\d\L
\left[\alpha(\L,\infty)-\frac{1}{2}\left(1+\frac{1}{\L}\right)\right]\right),
\eqno(6.28)
$$
with the natural choice $\L_0=1$ yielding $c_0=6.23367$.
The effective waiting time reads
$$
\s\eff\approx 2\s\left(1-\frac{2}{\L(\s)}+\cdots\right)
\approx 2\s\left(1-\frac{2}{\ln\s}+\cdots\right).
\eqno(6.29)
$$

\noindent ---
At finite (low) temperature and for long enough times,
the scaling function $\Phi$ blows up exponentially, as
$$
\Phi(\L(\s),\LL)\approx c_0 b(\LL)\e^{\s/t\eq\deux},
\eqno(6.30)
$$
with
$$
\s\eff=t\eq\deux\approx\frac{1}{B(\LL,\LL)}
=\frac{(\LL-1)\e^{\LL}}{\LL^2}\big(\LL^2\e^{-\LL}I(\LL)+1-\LL\big),
\eqno(6.31)
$$
in agreement with eq. (5.27).
The prefactor $b(\LL)$ can be given a convergent integral expression,
similar to eq. (6.13).

\noindent ---
In the crossover region,
when $\s$ is comparable to the relaxation times $t\eq\un$ and $t\eq\deux$,
the scaling function $\Phi(\L(\s),\LL)$ changes over
from the approximately square-root behaviour (6.27)
to the exponential behaviour (6.30).
The scaling behaviour in this crossover can be described,
to leading order in $1/\L$, by approximating eq. (6.21) as
$$
\alpha(\L,\LL)\approx\frac{1}{2\left(1-\e^{\L-\LL}\right)}.
\eqno(6.32)
$$
This equation can be integrated, to leading order in $1/\L(\s)$ and $1/\LL$.
We thus obtain
$$
\Phi(\L(\s),\LL)\approx\left(\e^{-\L(\s)}-\e^{-\LL}\right)^{-1/2},
\eqno(6.33)
$$
implying the estimate $\ln b(\LL)\approx\LL/2$ for $\LL$ large.
Accordingly, the effective waiting time departs from its linear growth (6.29)
in $\s$ to saturate to the value of the relaxation time $t\eq\deux$.
This quantity admits the simple expression
$$
\s\eff\approx\frac{2\e^{\L(\s)}}{\L(\s)},
\eqno(6.34)
$$
again to leading order in $1/\L(\s)$, throughout the crossover.

The above results allow us to describe the behaviour
of the correlation function in the $\alpha$ regime,
by inserting into the scaling law (6.23)
the estimates (6.27), (6.30), (6.33) for each of the times $\s$ and $t$.
The absolute prefactors $c_0$ cancel out,
so that the scaling law (6.23) does not depend on the initial conditions,
in agreement with the fact that only large times enter the expression (6.22).
In particular, in the aging regime $(\s$ and $t\ll t\eq\deux)$,
both $\Phi$ functions have the behaviour (6.27),
so that we obtain
$$
C(t,\s)\approx\left(\frac{\s}{t}\right)^{1/2}\frac{\ln\s}{\ln t},
\eqno(6.35)
$$
recovering the result of the zero-temperature analysis [11],
recalled in eq. (1.1).
In the converse situation $(\s,t\gg t\eq\deux)$,
both $\Phi$ functions have the exponential behaviour (6.30),
so that the equilibrium correlation function (5.31) is recovered.

\medskip
\noindent$\bullet$ {\it Energy response function}

The response function $r(t,\s)$ is expected to behave differently
from the correlation function $c(t,\s)$,
and especially to exhibit an appreciable variation in the $\beta$ regime
$(\theta=t-\s\sim 1)$,
in analogy with its equilibrium behaviour, described in Section 5.

The variation of $r(t,\s)$ throughout the nonequilibrium $\beta$ regime
can be determined by considering $h_0(t,\s)$ as a function of $\theta$,
and by solving eq. (3.9c) accordingly, by means of a Laplace transform,
as in Section 5, up to the following differences:
take $\L$ equal to $\L(\s)$, and not to its equilibrium value $\LL$;
consider $f_0(t)$ and $f_1(t)$ as variable quantities,
given by the dynamical equations (2.5).
With these assumptions, closed equations for the Laplace transforms
$\hat h_0(p)$, $\hat h_1(p)$ and $\hat Y_h(p)$ can be obtained,
yielding after some algebra
$$
r(t,\s)\approx\mu(\s)f_0(\s)R(\L(\s),\theta),
\eqno(6.36)
$$
with
$$
\mu(\s)f_0(\s)\approx
\frac{(\L(\s)-1)\left(\L(\s)^2\e^{-\beta}+A(\L(\s),\LL)\right)}{\L(\s)^2},
\eqno(6.37)
$$
and where $R(\L(\s),\theta)$ is obtained by replacing $\LL$ by $\L(\s)$
in the expression of the normalised response function
$R\eq(\theta)=\e^{\LL}r\eq(\theta)$ at equilibrium,
with $r\eq(\theta)$ being given in eq. (5.25a).
The result (6.36) holds at low temperature, and for $\s\gg 1$
and $\theta$ finite.

The response function exhibits a $\beta$-to-$\alpha$ plateau
for $1\ll\theta\ll t\eq\deux$, with a plateau value
$$
r\pl(\s)\approx\mu(\s)f_0(\s)R\pl(\L(\s)),
\eqno(6.38)
$$
with (see eqs. (5.30), (5.36))
$$
R\pl(\L(\s))\approx\frac{1}{\L(\s)^2\e^{-\L(\s)}I(\L(\s))+1-\L(\s)}.
\eqno(6.39)
$$
The result (6.38) interpolates between the equilibrium value
$r\pl(\s)=a_1$ for $\s\gg t\eq\deux$ and the zero-temperature value
$$
r\pl(\s)\approx\frac{\L(\s)-1}
{\L(\s)^2I(\L(\s))\big(\L(\s)^2\e^{-\L(\s)}I(\L(\s))+1-\L(\s)\big)}
\eqno(6.40)
$$
for $\s\ll t\eq\deux$.
The plateau value (6.38) is the appropriate initial condition
to be inserted into eq. (6.18), in order to describe the nonequilibrium
$\alpha$ relaxation of the response function.
We thus obtain
$$
r(t,\s)\approx r\pl(\s)C(t,\s).
\eqno(6.41)
$$
The above discussion of the scaling behaviour of the correlation function
$C(t,\s)$ applies to the response function
$r(t,\s)$ throughout the low-temperature $\alpha$ regime.

Figure 4 shows log-log plots of
the normalised correlation function and response function,
for various values of the inverse temperature $\beta$.
The data again nicely agree with the predictions (6.23), (6.41)
of the low-temperature analysis in the $\alpha$ regime.

\vfill\eject
%\medskip
\noindent$\bullet$ {\it Fluctuation-dissipation ratio}

The above results yield the following description
of the nonequilibrium behaviour of the \fd ratio $X(t,\s)$.
In the $\beta$ regime, we can approximate $g_0(t,\s)$ by unity.
Setting again $\theta=t-\s$, and using eq. (6.36), we obtain
$$
X(t,\s)\approx\frac
{\L(\s)^3\mu(\s)f_0(\s)R(\L(\s),\theta)}
{\L(\s)^3\mu(\s)f_0(\s)R(\L(\s),\theta)+A(\L(\s),\LL)}.
\eqno(6.42)
$$
Hence $X(t,\s)$ also exhibits a $\beta$-to-$\alpha$ plateau
for $1\ll\theta\ll t\eq\deux$, with a plateau value
$$
X\pl(\s)\approx\frac
{\L(\s)^3\mu(\s)f_0(\s)R\pl(\L(\s))}
{\L(\s)^3\mu(\s)f_0(\s)R\pl(\L(\s))+A(\L(\s),\LL)}.
\eqno(6.43)
$$
This result interpolates between the equilibrium value
$X\pl(\s)=1$ for $\s\gg t\eq\deux$ and the value
$$
X\pl(\s)
\approx\frac{\L(\s)(\L(\s)-1)}{\L(\s)^2\e^{-\L(\s)}I(\L(\s))+(\L(\s)-1)^2}
%\approx 1-\frac{2}{\L(\s)^2}-\frac{4}{\L(\s)^3}+\cdots
\approx 1-\frac{2}{\L(\s)^2}+\cdots
\approx 1-\frac{2}{(\ln s)^2}+\cdots
\eqno(6.44)
$$
at zero temperature, and more generally for $\s\ll t\eq\deux$.

Finally, the evolution equations (6.18), (6.20) imply
that the \fd ratio $X(t,\s)$ remains equal to $X\pl(\s)$
for all subsequent times $(\theta=t-\s\gg 1)$.
In other words, the evolution of the \fd ratio
does not couple at all to the slow dynamics of the model
in the $\alpha$ regime.
Figure 5 illustrates this behaviour.

\vskip 14pt plus 2pt
\noindent{\bf 7 Discussion}
\vskip 12pt plus 2pt

The present work is a continuation of previous studies [9, 11] of the so-called
Backgammon model, introduced in the context of the dynamics of glasses [7].
We have focused our attention on the mean energy of the model,
its local (or diagonal, see below) correlation and response functions
at finite temperature, and the associated \fd ratio.
The analysis of equilibrium properties (Section 5)
has revealed a remarkable feature of the model,
namely the possibility of separating in a controlled way
the slow modes ($\alpha$ relaxation) from the fast ones ($\beta$ relaxation).
This property is at the basis of the analytical treatment of the dynamics
(Section 6), which becomes asymptotically exact
in the regime of nonequilibrium $\alpha$ relaxation:
low temperatures and long times ($\s$ and $t-\s\gg 1$).
This approach consists in eliminating the fast degrees of freedom,
corresponding to $\beta$ relaxation.
It allows a quantitative description of the aging regime
(correlation and response decay as a power law,
the effective waiting time grows linearly),
of the convergence to equilibrium
(correlation and response decay exponentially with
characteristic time $t\eq\deux$ of the equilibrium situation,
the effective waiting time saturates to $t\eq\deux$),
and of the crossover behaviour between them.
The present method is an extension of the approach developed
at zero temperature [11],
interpreted there as an adiabatic approximation.
It is worth noticing that the physical picture of $\alpha$
and $\beta$ relaxation was absent from this zero-temperature analysis.

The multiplicative scaling law (6.23) of the energy correlation function
in the $\alpha$ regime is one of the main results of this work.
It is consistent with the description of nonequilibrium glassy dynamics given
e.g. in ref. [6].
Especially, for any three (long enough) times,
we have $C(t_1,t_3)=F\big(C(t_1,t_2),C(t_2,t_3)\big)$, with $F(u,v)=uv$.
The associated fixed-point equation $u=F(u,u)$ only has
the trivial solutions $u=0$ and $u=1$.
This simplistic behaviour reflects that the correlation function
has no non-trivial plateau value.
An investigation of the density correlation and response functions,
and of the associated \fd ratio, will be the subject of a separate paper [13].
The density correlation function does exhibit a non-trivial plateau value [12].

In this work we concentrated on the diagonal parts of the
correlation and response functions of energy,
which involve the same box at two different times $\s$ and $t$.
These diagonal contributions are known to be the leading ones
in disordered systems, such as spin glasses.
In the present model, however, non-diagonal correlations,
involving say box number 1 at time $\s$
and box number 2 at time $t$, are not negligible in general.
Let us take the example of equal-time equilibrium correlations of the energy.
The diagonal correlation
$\langle E_1^2\rangle-\langle E_1\rangle^2=(f_0)\eq(1-(f_0)\eq)$
yields a contribution $C\eq^{\rm diag}=\beta^2(f_0)\eq(1-(f_0)\eq)\approx\beta$
to the low-temperature specific heat per box.
On the other hand, the total specific heat at low temperature,
given in eq. (C.8), reads $C\eq\approx 1$.
We have therefore $C\eq\ll C\eq^{\rm diag}$.
This demonstrates the role of non-diagonal correlations,
which screen the diagonal ones almost perfectly in this example.

The present work can also be extended in another direction.
In ref. [9] three variants of the Backgammon model,
with different dynamical rules,
were introduced and investigated at zero temperature.
The model studied in the present work is model A,
the other two being models B and C.
We wish to emphasise that only model A has the property
that the spectrum of relaxation times at equilibrium
can be split in a controlled way into an $\alpha$ and a $\beta$ component.
We also recall that only model A possesses entropic traps and a very slow
zero-temperature dynamics $(\L(t)\sim\ln t)$,
while $\L(t)$ obeys a power-law growth for models B and C [9].
These two properties therefore seem to go hand in hand.
Finally, we would like to mention a generalisation of model B
which undergoes a condensation transition at finite temperature [15].

\vfill\eject
%\vskip 14pt plus 2pt
\noindent{\bf Acknowledgements}
\vskip 12pt plus 2pt

It is a pleasure to thank S. Franz and F. Ritort
for interesting discussions during the elaboration of this work,
and J.P. Bouchaud for useful comments.

\vfill\eject
\noindent{\bf Appendix A. Derivation of dynamical equations}
\vskip 12pt plus 2pt

\noindent$\bullet$ {\it The case of the $f_k$, $g_k$, and $h_k$}

The rules defining the dynamics of the model are as follows.
For a system of $M$ boxes and $N$ particles,
during every time step $\d t=1/N$:

\noindent ---
a particle is drawn at random, which determines a departure box $d$.
The box $d$ is thus chosen with probability $\pi_d=N_d/N$,
since it contains $N_d$ particles;

\noindent ---
an arrival box $a$, different from $d$, is drawn at random
with uniform probability $\pi_a=1/(M-1)$;

\noindent ---
the move is accepted according to the Metropolis rule.

A configuration of the system is given by the occupation numbers of the boxes:
$\C=\{N_1,$ $N_2,$ $\ldots,$ $N_M\}$, with $\sum_iN_i=N$.
We first determine the transition rates appearing in the master equation
$$
\frac{\d}{\d t}P(\C)
=\sum_{\C'\ne\C}\M(\C\mid\C')P(\C')-\sum_{\C''\ne\C}\M(\C''\mid\C)P(\C).
\eqno({\rm A}.1)
$$
Let us consider the rate $\M(\C''\mid\C)$ for definiteness.
We use the following notation for the initial configuration:
$\C=\{N_1,\ldots,N_d,\ldots,N_a,\ldots,N_M\}=\{N_d,N_a\}$.
Then the final configuration is $\C''=\{N_d-1, N_a+1\}$.
The Metropolis acceptance rate reads
$$
W(N_d-1,N_a+1\mid N_d,N_a)=\W(\Delta S)=\min(1,\e^{-\Delta S}),
\eqno({\rm A}.2)
$$
where the change in action is
$$
\Delta S=S(N_d-1,N_a+1)-S(N_d,N_a)=-\beta(\delta_{N_d,1}-\delta_{N_a,0}).
\eqno({\rm A}.3)
$$
The possible transitions to be considered are
$$
\matrix{
\C=\{N_d>1, N_a>0\},\hfill&\Delta S=0,\hfill&
W(N_d-1,N_a+1\mid N_d,N_a)=1,\hfill\cr
\C=\{N_d>1, N_a=0\},\hfill&\Delta S=\beta,\hfill&
W(N_d-1,N_a+1\mid N_d,N_a)=\e^{-\beta},\hfill\cr
\C=\{N_d=1, N_a>0\},\hfill&\Delta S=-\beta,\hfill&
W(N_d-1,N_a+1\mid N_d,N_a)=1,\hfill\cr
\C=\{N_d=1, N_a=0\},\hfill&\Delta S=0,\hfill&
W(N_d-1,N_a+1\mid N_d,N_a)=1.\hfill\cr
}
\eqno({\rm A}.4)
$$
The transition rate $\M(\C''\mid\C)$ is thus given by
$$
\M(\C''\mid\C)=\M(N_d-1,N_a+1\mid N_d,N_a)=
\frac{1}{\d t}\pi_d\pi_a W(N_d-1,N_a+1\mid N_d,N_a).
\eqno({\rm A}.5)
$$

We now consider the reduced configuration
characterised by the occupation number of box number 1 alone:
$\C=\{N_1(t)\}=\{k\}$. The transition rates $\M(\C''\mid\C)$=$\M(k\pm 1\mid k)$
respectively correspond to the case where box number 1 is the arrival box
$(a=1$: upper sign) and the departure box $(d=1$: lower sign).
They are obtained by summing eq. (A.5) over all the boxes.
In the thermodynamic limit ($M,N\to\infty$, with a fixed density $\rho=N/M$),
we get\footnote{$\dagger$}{In this limit, it does not matter whether the boxes
which are summed upon are assumed to be different or not from box number 1.}
$$
\eqalign{
\M(k+1\mid k)&=\frac{1}{\d t}\sum_d\pi_d\pi_1
W(N_d-1,N_1+1\mid N_d,N_1)=\rho{\hskip 2.9cm}(k\ge1),\cr
\M(k-1\mid k)&=\frac{1}{\d t}\sum_a\pi_1\pi_a
W(N_1-1,N_a+1\mid N_1,N_a)=k(1-f_0+\e^{-\beta}f_0)\quad(k\ge2),\cr
\M(1\mid 0)&=\frac{1}{\d t}\sum_d\pi_d\pi_1
W(N_d-1,N_1+1\mid N_d,N_1)=f_1+\e^{-\beta}(\rho-f_1),\cr
\M(0\mid 1)&=\frac{1}{\d t}\sum_a\pi_1\pi_a
W(N_1-1,N_a+1\mid N_1,N_a)=1.\cr
}
\eqno({\rm A}.6)
$$
The rates $\M(\C\mid\C')$=$\M(k\mid k\pm 1)$ are also given by
these expressions.

With the notation $\M_{k\ell}[f_0(t),f_1(t)]=\M(k\mid\ell)$,
and setting $\rho=1$, we obtain the dynamical equations (2.5), (2.7), i.e.,
$$
\frac{\d f_k(t)}{\d t}
=\sum_{\ell\ge0}\M_{k\ell}\left[f_0(t),f_1(t)\right]f_\ell(t).
\eqno({\rm A}.7)
$$

The $g_k(t,\s)$ obey the similar equation (2.14),
since the time variable $\s$ plays the role of a parameter in the dynamics.
Finally, the $h_k(t,\s)$ obey eq. (2.26), as explained in Section 2.

\medskip
\noindent$\bullet$ {\it The case of the $\h_k$}

The only change with respect to the previous case
comes from the fact that the local inverse temperature of box number
1 is now $\beta_1(t)$.
The action therefore reads
$$
S=\beta_1(t)E_1+\beta\sum_{i=2}^M E_i
=-\beta_1(t)\delta_{N_1,0}-\beta\sum_{i=2}^M\delta_{N_i,0},
\eqno({\rm A}.8)
$$
hence
$$
\Delta S=S(N_d-1,N_a+1)-S(N_d,N_a)=\beta_a\delta_{N_a,0}-\beta_d\delta_{N_d,1}.
\eqno({\rm A}.9)
$$
The possible transitions to be considered are
$$
\matrix{
\C=\{N_d>1,N_a>0\},\hfill&\Delta S=0,\hfill&
W(N_d-1,N_a+1\mid N_d,N_a)=1,\hfill\cr
\C=\{N_d>1,N_a=0\},\hfill&\Delta S=\beta_a,\hfill&
W(N_d-1,N_a+1\mid N_d,N_a)=\e^{-\beta_a},\hfill\cr
\C=\{N_d=1,N_a>0\},\hfill&\Delta S=-\beta_d,\hfill&
W(N_d-1,N_a+1\mid N_d,N_a)=1,\hfill\cr
\C=\{N_d=1,N_a=0\},\hfill&\Delta S=\beta_a-\beta_d,\hfill&
W(N_d-1,N_a+1\mid N_d,N_a)=\W(\Delta S).\hfill\cr
}
\eqno({\rm A}.10)
$$
Pursuing the derivation as above, we obtain the dynamical equations (2.19).

\vfill\eject
\noindent{\bf Appendix B. Solution of eqs. (3.2)
by the method of characteristics}
\vskip 12pt plus 2pt

We recall how to solve, using the method of characteristics,
a partial differential equation of the form
$$
\p{}{t}G(x,t,\s)=
(x-1)\left(G(x,t,\s)-{1\over\L(t)}\p{}{x}G(x,t,\s)-Y(t,\s)\right)
\eqno({\rm B}.1)
$$
for $t>s$, where $Y(t,\s)$ is a given function,
and where the initial value $G(x,\s,\s)$ is also given.

The starting point of the method consists in writing the proportionality
$$
\d t=\frac{\d x}{(x-1)/\L(t)}=\frac{\d G}{(x-1)(G-Y)}.
\eqno({\rm B}.2)
$$

\noindent
The first equality in eq. (B.2) yields
$$
x-1=y\e^{\tau(t)},
\eqno({\rm B}.3)
$$
with
$$
\tau(t)=\int_0^t\frac{\d u}{\L(u)},
\eqno({\rm B}.4)
$$
and where the constant $y$ parametrises the characteristic curves.

\noindent
The second equality in eq. (B.2) then reads
$$
\frac{\d G}{\d t}=y\e^{\tau(t)}(G-Y),
\eqno({\rm B}.5)
$$
which can be integrated by varying the constant.
This yields, taking in account the initial condition,
$$
G(x,t,\s)=\e^{(x-1)D(t,\s)}G\left(1+(x-1)\e^{\t(\s)-\t(t)},\s,\s\right)
+\int_\s^t\d u\,\K(x,t,u)Y(u,\s),
\eqno({\rm B}.6)
$$
with the definitions
$$
\eqalign{
D(t,u)&=\int_u^t\d v\,\e^{\t(v)-\t(t)},\cr
\K(x,t,u)&=(1-x)\e^{\t(u)-\t(t)+(x-1)D(t,u)}.
}
\eqno({\rm B}.7)
$$

\vfill\eject
\noindent{\bf Appendix C. Equilibrium thermodynamics}
\vskip 12pt plus 2pt

Consider a system made of $M$ boxes containing $N$ particles.
As in Appendix A, we define a configuration $\C$ of the system as given by the
occupation numbers of the boxes:
$\C=\{N_1,$ $N_2,$ $\ldots,$ $N_M\}$, with $\sum_iN_i=N$.
The partition function of the system reads
$$
Z_{M,N}=\frac{1}{N!}\sum_{\big\{N_i\ge0,\sum_iN_i=N\big\}}{N!\over\prod_iN_i!}
\,\e^{\beta\sum_i\delta_{N_i,0}}.
\eqno({\rm C}.1)
$$
The combinatorial weight inside the sum represents the number of microstates
corresponding to each configuration $\C$,
where a microstate is defined by assigning in which box lies
each of the $N$ particles.
The overall $1/N!$ in front of the sum takes account of the fact that a global
relabeling of the identical particles does not affect equilibrium properties.
This statistics, used e.g. for ideal gases of classical particles, is
usually referred to as Maxwell-Boltzmann statistics [16].

Inserting into eq. (C.1) the contour integral representation
${\displaystyle\oint}\frac{\d z}{2\pi iz}\,z^{\sum_iN_i-N}$ for the constraint,
we obtain
$$
Z_{M,N}=\oint\frac{\d z}{2\pi iz}\,s(z)^M\,z^{-N},
\eqno({\rm C}.2)
$$
with
$$
s(z)=\sum_{k\ge 0}\frac{\e^{\beta\delta_{k,0}}}{k!}z^k=\e^z+\e^\beta-1.
\eqno({\rm C}.3)
$$
We have in particular $Z_{M,N}=M^N/N!$ at infinite temperature.

In the thermodynamic limit $(M,N\to\infty)$, the density $\rho=N/M$ being
fixed, the method of steepest descent can be applied to the integral (C.2).
The saddle-point equation reads $\d s(z)/\d z=\rho s(z)/z$.
The saddle-point value $z_c$ of $z$,
which is by definition the thermodynamical fugacity of the model,
is thus related to temperature and density through
$$
\e^\beta=1+(z_c/\rho-1)\e^{z_c}.
\eqno({\rm C}.4)
$$
The free energy of the model is an extensive quantity,
and its value per box reads
$$
-\beta F=\ln s(z_c)-\rho\ln z_c=z_c+(1-\rho)\ln z_c-\ln\rho.
\eqno({\rm C}.5)
$$

Physical quantities can be derived along the same lines.
For instance, by restricting the sum in eq. (C.1)
to the configurations $\{N_i\}$ such that $\sum_iN_i=N$ and $N_1=k$,
we obtain for a finite system
$$
(f_k)_{M,N}=\frac{1}{Z_{M,N}}\frac{\e^{\beta\delta_{k,0}}}{k!}
\oint\frac{\d z}{2\pi iz}\,s(z)^{M-1}\,z^{k-N},
\eqno({\rm C}.6)
$$
and in the thermodynamic limit
$$
(f_k)\eq=\rho\exp\left(-z_c+\beta\delta_{k,0}\right)\frac{z_c^{k-1}}{k!}.
\eqno({\rm C}.7)
$$

We now restrict ourselves to the case $\rho=1$,
considered in the body of the paper.
Eqs. (5.2) and (C.4) show that the fugacity $z_c$
is identical to the equilibrium value $\LL$ of $\L(t)$,
while eq. (5.1) for the occupation probabilities coincides with eq. (C.7).

The low-temperature behaviour of various thermodynamical quantities,
such as the fugacity $\LL=z_c$, the internal energy $E\eq=-(f_0)\eq$,
and the specific heat $C\eq=-\beta^2\,\d E\eq/\d\beta$,
is obtained by expanding the above results for $\beta\to\infty$:
$$
\eqalign{
\LL&=\beta-\ln\beta+\frac{\ln\beta+1}{\beta}
+\frac{\ln^2\beta-1}{2\beta^2}+\cdots,\cr
E\eq&=-1+\frac{1}{\beta}+\frac{\ln\beta}{\beta^2}
+\frac{\ln^2\beta-\ln\beta-1}{\beta^3}+\cdots,\cr
C\eq&=1+\frac{2\ln\beta-1}{\beta}
+\frac{3\ln^2\beta-5\ln\beta-2}{\beta^2}+\cdots
}
\eqno({\rm C}.8)
$$
The constant equilibrium specific heat $C\eq\approx1$ at low temperature
is a somewhat pathological feature of the model [16],
that is shared by several classical systems,
such as the ideal gas and the harmonic oscillator.

\vfill\eject
{\parindent 0em
{\bf References}
\vskip 12pt plus 2pt

[1] J. Langer, in {\it Solids far from Equilibrium}, C. Godr\`eche, ed.
(Cambridge University Press, 1991).

[2] A.J. Bray, Adv. Phys. {\bf 43} (1994), 357.

[3] {\it Nonequilibrium Statistical Mechanics in One Dimension}, V. Privman,
ed. (Cambridge University Press, 1997).

[4] {\it Heidelberg Colloquium on Glassy Dynamics}, J.L. van Hemmen and I.
Morgenstern, eds. (Springer Verlag, 1987).

[5] {\it Frontiers in Materials Science} (5 articles on glasses and amorphous
materials), Science {\bf 267} (1995), 1924--1953.

[6] For a recent review, see: J.P. Bouchaud, L.F. Cugliandolo, J. Kurchan,
and M. M\'ezard, cond-mat/9702070, and references therein.

[7] F. Ritort, Phys. Rev. Lett. {\bf 75} (1995), 1190.

[8] S. Franz and F. Ritort, Europhys. Lett. {\bf 31} (1995), 507.

[9] C. Godr\`eche, J.P. Bouchaud, and M. M\'ezard, J. Phys. A {\bf 28}
(1995), L 603.

[10] S. Franz and F. Ritort, J. Stat. Phys. {\bf 85} (1996), 131.

[11] C. Godr\`eche and J.M. Luck, J. Phys. A {\bf 29} (1996), 1915.

[12] S. Franz and F. Ritort, J. Phys. A {\bf 30} (1997), L 359.

[13] S. Franz, C. Godr\`eche, J.M. Luck, and F. Ritort, in preparation.

[14] L.F. Cugliandolo, J. Kurchan, and L. Peliti, cond-mat/9611044.

[15] P. Bialas, Z. Burda, and D. Johnston, cond-mat/9609264.

[16] B.J. Kim, G.S. Jeon, and M.Y. Choi, Phys. Rev. Lett. {\bf 76} (1996),
4648.

\vfill\eject
\noindent{\bf Figure Captions}
\vskip 12pt plus 2pt

{\bf Figure 1:}
Logarithmic plot of the relaxation times $t\eq\un$ (mean energy)
and $t\eq\deux$ (energy correlation and response functions),
against inverse temperature $\beta$.
Full lines: exact relaxation times (5.13), (5.24).
Dashed lines: low-temperature estimates (5.17), (5.28).

\smallskip
{\bf Figure 2:}
Log-log plot of (a) the normalised correlation function $C\eq(\theta)$
and (b) the normalised response function $R\eq(\theta)$ at equilibrium,
as given by the exact expressions (5.25),
against the time difference $\theta=t-\s$,
for various values of inverse temperature $\beta$, indicated on the curves.

\smallskip
{\bf Figure 3:}
Plot of $\L(t)$ against $\ln t$,
for various values of inverse temperature $\beta$, indicated on the curves.
Full lines: exact values,
obtained by numerical integration of the dynamical equations (2.5).
Dashed lines: low-temperature estimates (6.6).

\smallskip
{\bf Figure 4:}
Log-log plot of (a) the normalised correlation function $C(t,\s)$
and (b) the normalised response function $R(t,\s)$,
against the time difference $t-\s$, for a waiting time $\s=100$,
and various values of inverse temperature $\beta$, indicated on the curves.
Full lines: exact values,
obtained by numerical integration of the dynamical equations (2.14), (2.26).
Dashed lines: predictions (6.23), (6.41) of the low-temperature analysis
in the $\alpha$ regime.

\smallskip
{\bf Figure 5:}
Plot of the \fd ratio $X(t,\s)$ against the time difference $t-\s$,
for a waiting time $\s=100$,
and various values of inverse temperature $\beta$, indicated on the curves.
Full lines: exact values,
obtained by numerical integration of the dynamical equations (2.14), (2.26).
Arrows: low-temperature estimates (6.43) for the plateau value $X\pl$.
}
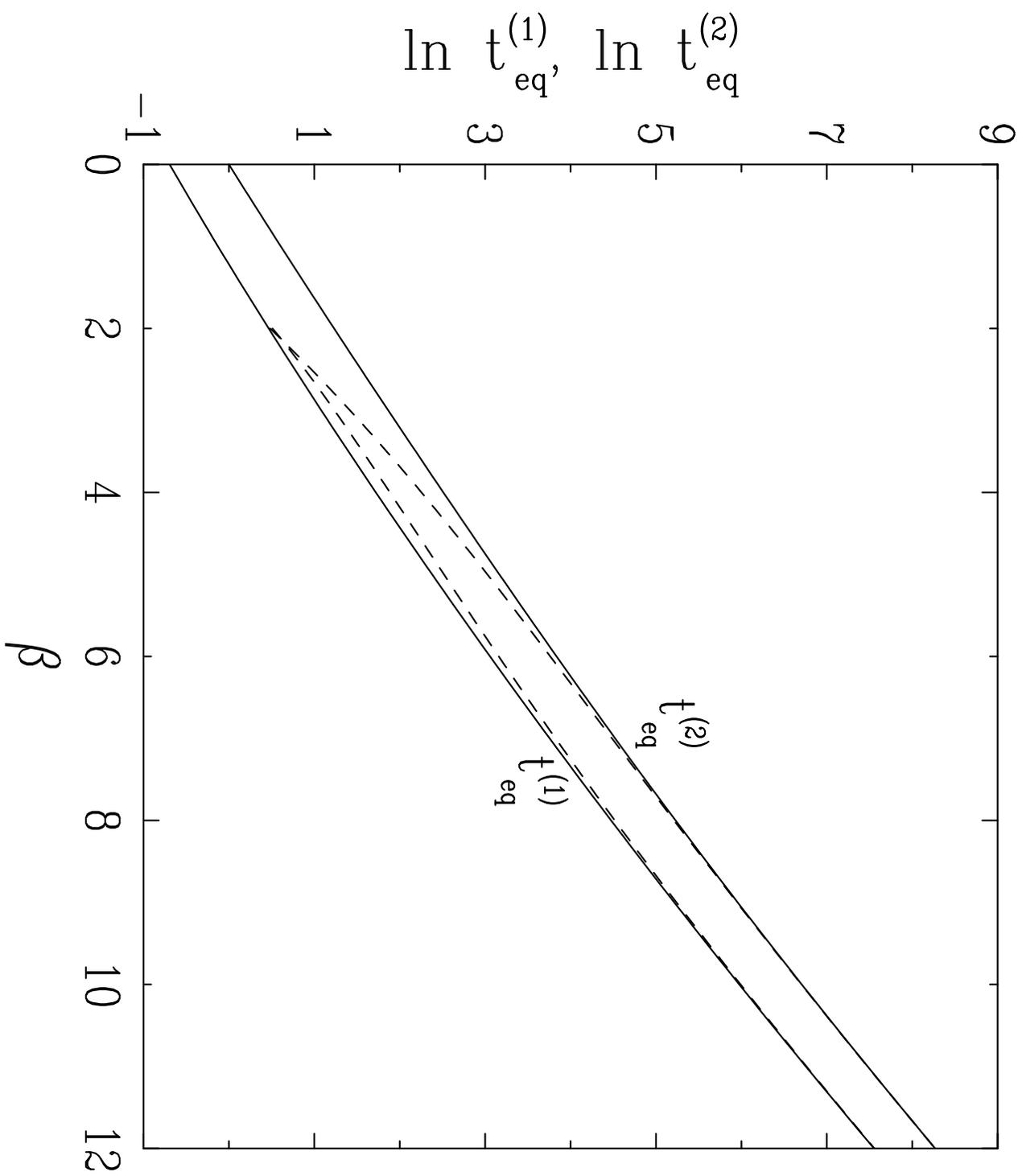
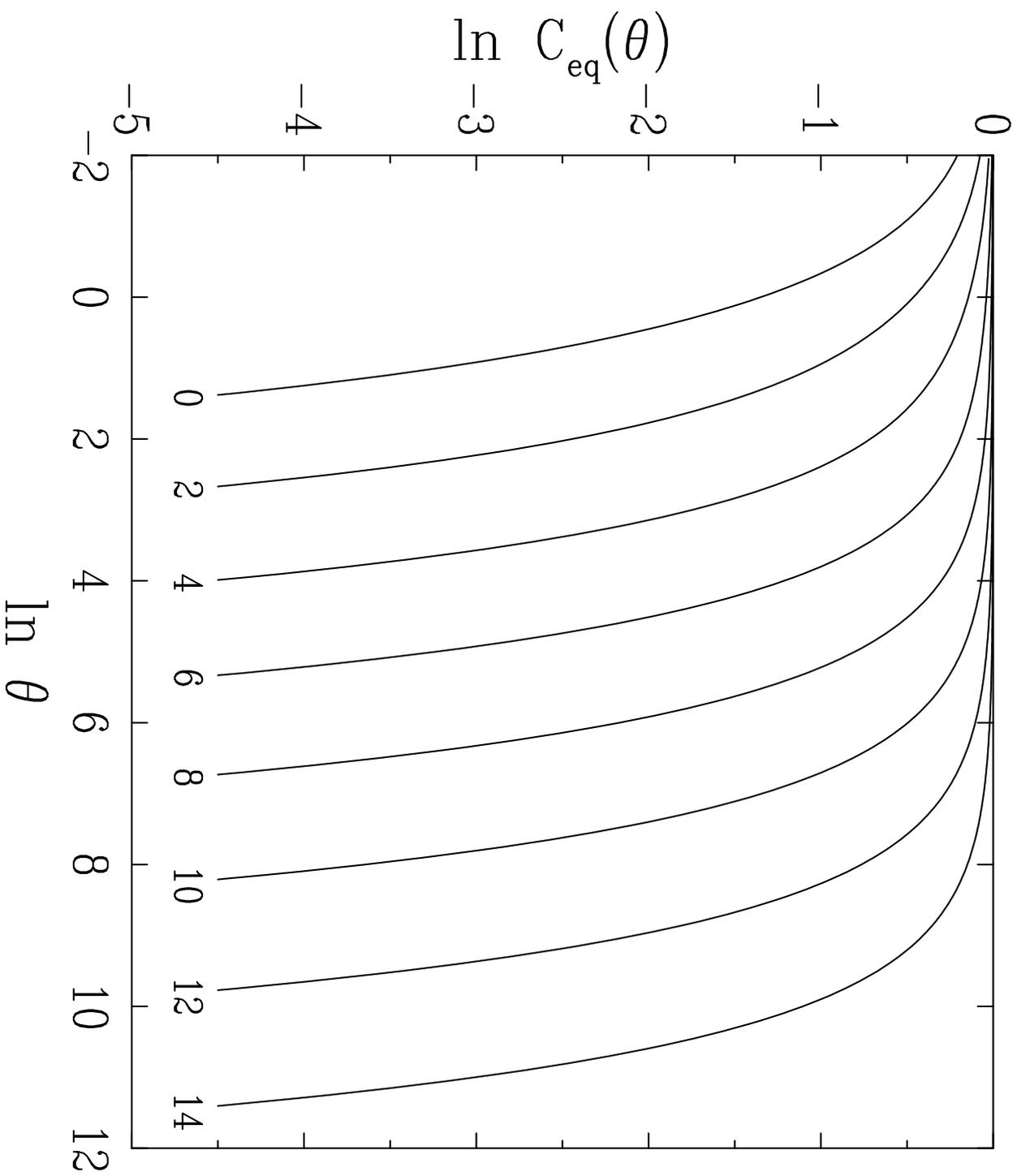
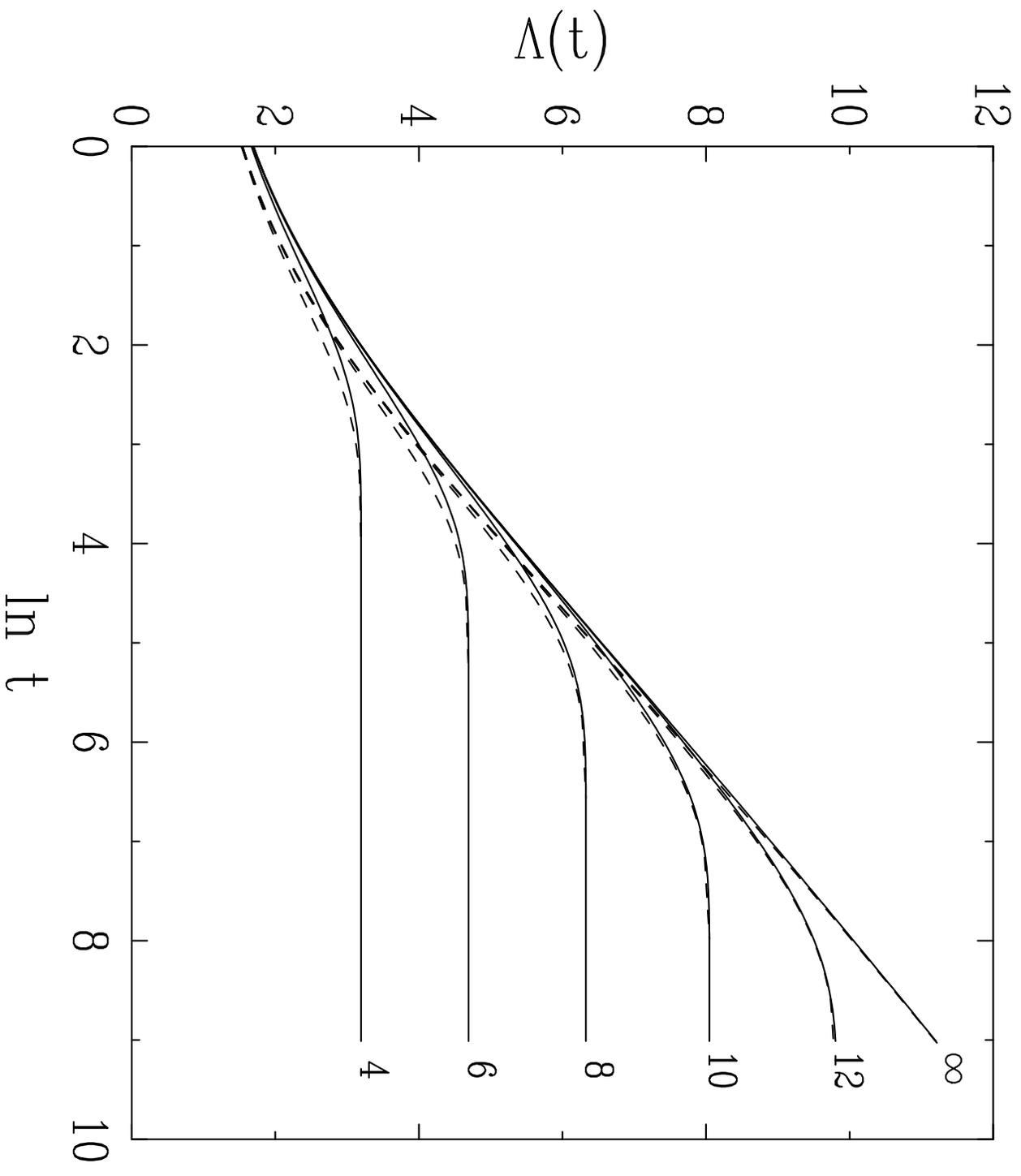
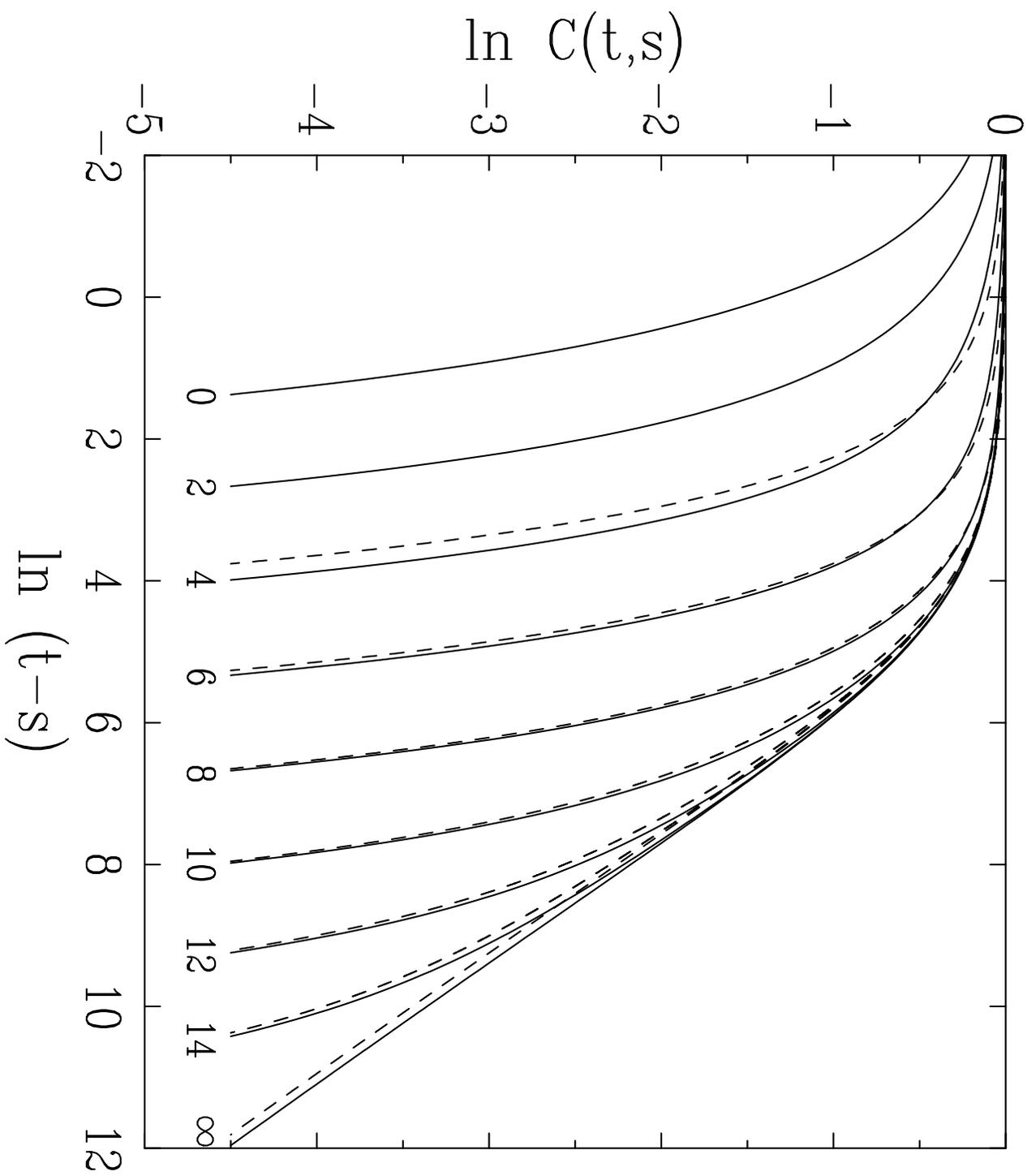
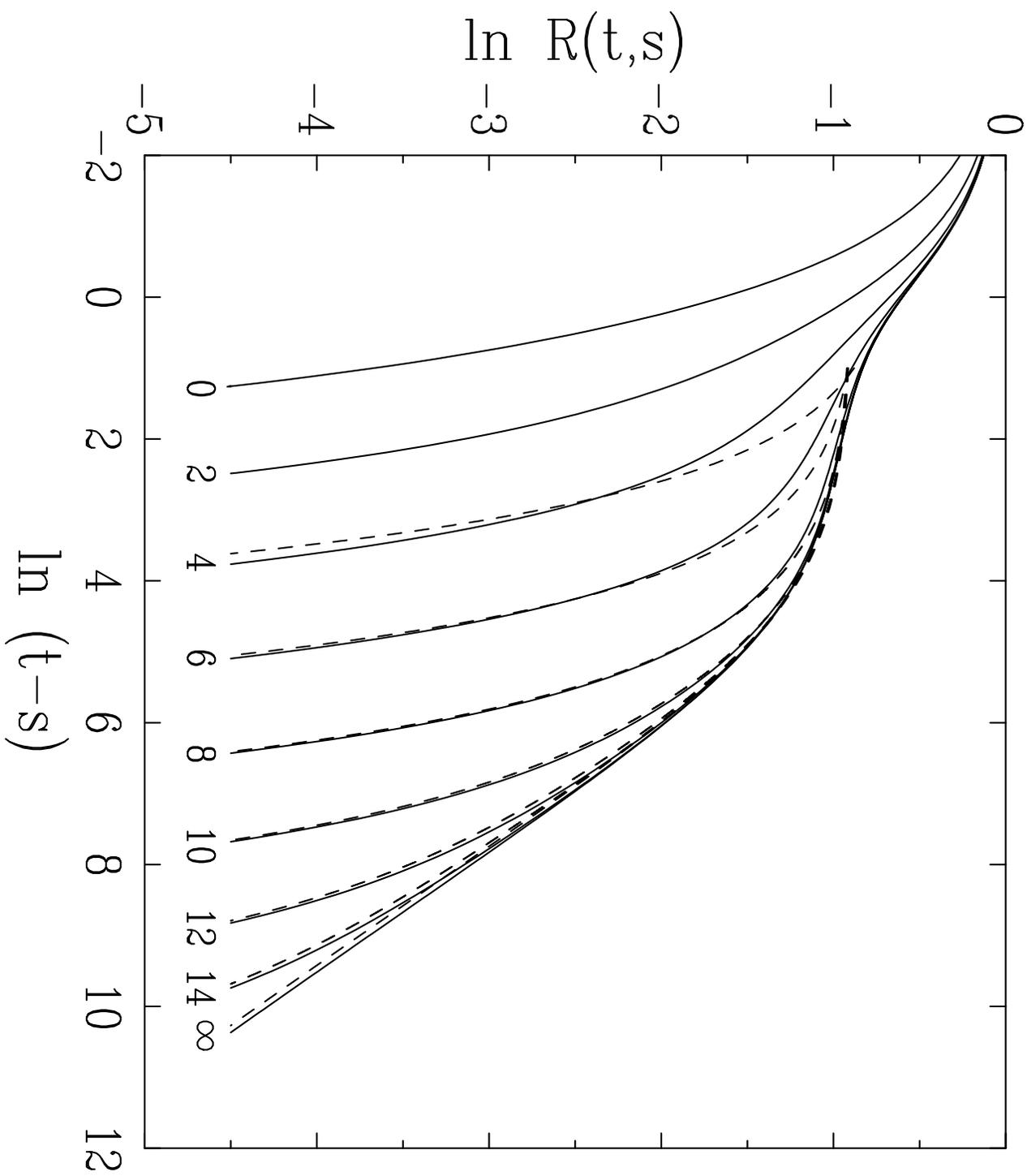
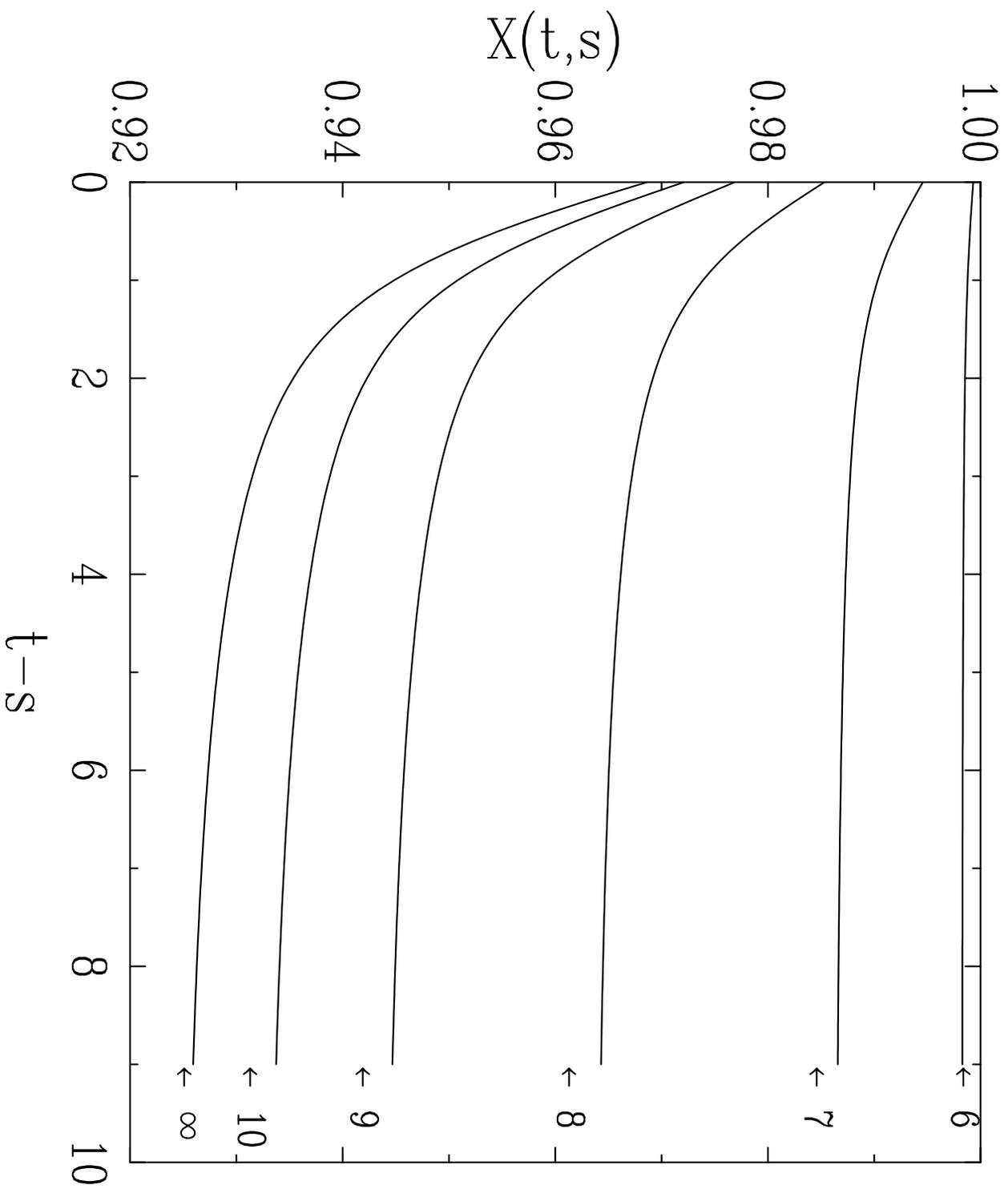
\bye